\documentclass[12pt]{article}
\pdfoutput=1

\usepackage{amsmath}
\usepackage{amssymb}
\usepackage{epsfig}
\usepackage{graphicx}
\usepackage{cite}
\usepackage{multirow}
\usepackage{longtable}
\usepackage{lscape}
\usepackage{bbm}
\usepackage{dcolumn}
\usepackage{rotating}

\allowdisplaybreaks[4] 

\newcommand{\capdef}{}
\newcommand{\mycaption}[2][\capdef]{\renewcommand{\capdef}{#2}%
        \caption[#1]{{\footnotesize #2}}}
\makeatletter
\renewcommand{\fnum@table}{\textbf{\tablename~\thetable}}
\renewcommand{\fnum@figure}{\textbf{\figurename~\thefigure}}
\makeatother

\newcounter{myenumi}

\renewcommand{\themyenumi}{\roman{myenumi}}
{\end{list}}

\newlength{\myem}
\newcounter{mysubequation}[equation]

\makeatletter
\renewcommand{\section}{\@startsection{section}{1}{0em}{-\baselineskip}%
{\baselineskip}{\normalfont\large\bfseries}}
\renewcommand{\subsection}%
{\@startsection{subsection}{2}{0em}{-0.7\baselineskip}%
{0.7\baselineskip}{\normalfont\bfseries}}
\makeatother

\newcommand{\bi}{\begin{itemize}}
\newcommand{\ei}{\end{itemize}}

\newcommand{\be}{\begin{equation}}
\newcommand{\ee}{\end{equation}}
\newcommand{\bea}{\begin{eqnarray}}
\newcommand{\eea}{\end{eqnarray}}

\newcommand{\xhat}{{\widehat X}}
\newcommand{\rhat}{{\widehat R}}

\newcommand{\ie}{{\it i.e.}}

\newcommand{\eg}{{\it e.g.}}

\newcommand{\cf}{{\it cf.}}

\newcommand{\eq}{Eq.}

\newcommand{\fig}{Fig.}

\newcommand{\Ref}{Ref.}
\newcommand{\Refs}{Refs.}
\newcommand{\Sec}{Sec.}

\newcommand{\App}{Appendix}

\newcommand{\Tab}{Table}

\newcommand{\equ}[1]{\eq~(\ref{equ:#1})}
\newcommand{\figu}[1]{\fig~\ref{fig:#1}}

\begin{document}

\begin{titlepage}

\renewcommand{\thefootnote}{\alph{footnote}}

\vspace*{-3.cm}
\begin{flushright}

\end{flushright}


\renewcommand{\thefootnote}{\fnsymbol{footnote}}

{\begin{center} {\large\bf
Interplay of energy dependent astrophysical neutrino flavor ratios and new physics effects
}
\end{center}}

\renewcommand{\thefootnote}{\alph{footnote}}

\vspace*{.8cm}
\vspace*{.3cm}
{\begin{center} {\large{\sc
                Poonam Mehta\footnote[1]{\makebox[1.cm]{Email:}
                poonam@rri.res.in} and
                Walter Winter\footnote[2]{\makebox[1.cm]{Email:}
                winter@physik.uni-wuerzburg.de}
                }}
\end{center}}
\vspace*{0cm}
{\it
\begin{center}

\footnotemark[1]%
       Raman Research Institute, \\ C. V. Raman Avenue, Bangalore 560 080, India

\footnotemark[2]%
       Institut f{\"u}r Theoretische Physik und Astrophysik, \\ Universit{\"a}t W{\"u}rzburg,
       97074 W{\"u}rzburg, Germany

\end{center}}

\vspace*{1.5cm}

\begin{center}
{\Large \today}
\end{center}

{\Large \bf
\begin{center} Abstract \end{center}  }

 We discuss the importance of flavor ratio measurements in neutrino telescopes, such as by
measuring the ratio between muon tracks to cascades, for the purpose of extracting  new physics
  signals encountered by astrophysical neutrinos  during propagation from the source to the detector. The detected flavor
    ratios not only carry the energy information of specific new physics scenarios which alter the transition
    probabilities in distinctive ways, but also the energy dependent flavor composition at the source. In the
     present work, we discuss the interplay of these two energy dependent effects  and identify which new physics scenarios can be distinguished from the detected flavor ratios as a function of astrophysical parameters.
We use a recently developed self-consistent neutrino production model as our toy model to generate energy
dependent source flavor ratios and discuss (invisible) neutrino decay and quantum decoherence as specific new
physics examples.
   Furthermore, we identify potentially interesting classes of sources on
 the Hillas plot for the purpose of new physics searches. We find that sources
with substantial magnetic fields $10^3 \, \mathrm{Gauss} \lesssim B
 \lesssim 10^6 \, \mathrm{Gauss}$, such as  Active Galactic Nuclei (AGN) cores,
white dwarfs, or maybe  gamma-ray bursts, have, in principle, the best discrimination
 power for the considered new physics scenarios, whereas AGN jets, which typically perform
  as pion beam sources, can only discriminate few sub cases in the new physics effects. The
    optimal parameter region somewhat
depends on the class of new physics effect considered.

\vspace*{.5cm}

\end{titlepage}

\newpage

\renewcommand{\thefootnote}{\arabic{footnote}}
\section{Introduction}

Even though extraterrestrial high energy neutrino flux has not been detected yet,
  efforts are made to detect neutrinos from far beyond the Sun and the Supernova 1987A.
   Astrophysical neutrinos at high energies open up an entirely new window to infer the
   properties of  production sites as well as neutrinos themselves. The
  kilometer-scale neutrino telescopes such as IceCube~\cite{Ahrens:2002dv} and
   KM3NeT~\cite{1742-6596-203-1-012124} are designed to detect high energy ($E \gtrsim 10^{11}$ eV)
    neutrinos from various astrophysical sources.
   Among the candidate sources, the most prominent extragalactic ones are Active Galactic Nuclei
    (AGNs)~\cite{Stecker:1991vm,Mannheim:1993,Mucke:2000rn,Aharonian:2002} and
  Gamma-Ray Bursts (GRBs)~\cite{Waxman:1997ti}, see \Ref~\cite{Becker:2007sv} for a
   review and \Ref~\cite{Rachen:1998fd} for
     the general theory governing photohadronic neutrino production from astrophysical sources.
   There are generic theoretical  bounds for the diffuse neutrino flux from sources transparent as well as
    opaque to neutrons~\cite{Waxman:1998yy,Mannheim:1998wp}.
The diffuse flux bounds from these sources are being touched by IceCube, see, \eg, \Ref~\cite{IcGRB:2011qc} for GRBs. Apart from that, there is
also the possibility to detect point sources for which even the optical
       counterpart is blocked, the so-called ``hidden sources''~\cite{Razzaque:2009kq}.

We focus on the possibility to detect the flavor of astrophysical neutrinos.
 Although neutrino telescopes such as IceCube can not detect the flavor of the neutrinos directly, one can, in the simplest case,
 make use of the two kinds of event topologies: tracks (which are induced by muons) and cascades/showers
  (electromagnetic and hadronic, induced by $\nu_e$ and $\nu_\tau$, respectively) to construct an observable track to cascade flavor ratio as a flavor dependent quantity. Note that this flavor ratio is defined in the spirit of the very first search for cascades from extragalactic neutrino sources, which has been very recently conducted in \Ref~\cite{IcCascade:2011ui} for IceCube-22.
This flavor ratio allows us to infer particle physics properties of neutrinos~\cite{Pakvasa:1981ci,Learned:1994wg,Beacom:2003nh,Beacom:2003eu,Pakvasa:2008nx,Pakvasa:2010jj}, extract
 information on the flavor mixing
parameters~\cite{Farzan:2002ct,Beacom:2003zg,Serpico:2005sz,Serpico:2005bs,Bhattacharjee:2005nh,Winter:2006ce,Majumdar:2006px,Meloni:2006gv,Rodejohann:2006qq,Xing:2006xd,Pakvasa:2007dc,Blum:2007ie,Hwang:2007na,Choubey:2008di,Esmaili:2009dz} and identify such sources~\cite{Xing:2006uk,Choubey:2009jq}. Since the flux normalization drops out of this ratio, it is relatively robust with respect to astrophysical uncertainties. We will elaborate on this point further in \Sec~\ref{sec:flr}.

     The high energy neutrinos are conventionally expected to originate from the decay of charged pions
      (pion beam source) produced via photohadronic processes ($p\gamma$)  or inelastic
      ($pp$) collisions. From $\pi \to \mu \to e$ decay chain, the flavor composition at the source is
      given by  $\Phi_e^0:\Phi_\mu^0:\Phi_\tau^0 = 1:2:0$ ($\Phi_{\alpha}^0$ represents sum of neutrino
       and anti-neutrinos of a given flavor)~\footnote{At neutrino telescopes
        without charge identification capability,
        $p\gamma$ and $pp$ processes are indistinguishable since both lead to the same flavor
          composition $1:2:0$ (neutrinos and anti-neutrinos are added).   However, one may use
           the Glashow resonance ($E \simeq 6.3$ PeV) initiated by $\bar\nu_e$ to discriminate
           between the primary processes which differ in $\pi^+ - \pi^-$
            symmetry~\cite{Bhattacharjee:2005nh,Anchordoqui200518}.}. However, this picture is
             over-simplified and it was
          pointed out in \Refs~\cite{Rachen:1998fd,Kashti:2005qa} that energy losses in strong magnetic
            fields, which dominantly affect the muons for a pion beam source, changes
            $\Phi_e^0:\Phi_\mu^0:\Phi_\tau^0$ to $0:1:0$ at high energies (muon damped source). This established
             the energy dependence of the flavor composition of a given source and one can expect a
smooth transition from one type of source to another as a function of the neutrino
energy~\cite{Kachelriess:2006fi,Lipari:2007su,Kachelriess:2007tr}, mostly depending on the
 cooling processes of
the intermediate muons, pions and kaons (or even charmed mesons, which we do not consider). Other mechanisms, such as neutron decay,  define a new class of
sources, and even all special classes may be recovered as a function of
energy~\cite{Hummer:2010ai}. At low energies, neutron
           decays may dominate
           the flux, where the neutrons are generated in the photodissociation of heavy nuclei or
            photohadronic interactions, which leads to $1:0:0$ (neutron beam source)~\cite{Anchordoqui:2003vc}. If the cooled muons pile up at low energies, one may even have
$1:1:0$ from muon decays only (muon beam source)~\cite{Hummer:2010ai}. A simple-minded
               model for implementing the photohadronic interactions in cosmic accelerators
                was recently developed by H\"{u}mmer, Maltoni, Winter and
Yaguna [HMWY]~\cite{Hummer:2010ai}.
 In the HMWY model, charged pions are produced from
photohadronic ($p\gamma$) interactions  between protons and the synchrotron photons from
             co-accelerated electrons (positrons). The photohadronic interactions are computed using
            an efficient state-of-the-art method described in \Ref~\cite{Hummer:2010vx}, based
             on the physics of SOPHIA~\cite{Mucke:1999yb},  and the helicity-dependent muon decays
               from \Ref~\cite{Lipari:2007su} are included.
               The toy model relies on relatively few astrophysical parameters, the most
                important ones  being  the size of the acceleration region (R),   the magnetic field strength at the source ($B$),
  and  the injection index ($\alpha$) which is assumed to be universal for protons and electrons/positrons. In the HMWY
   model, the  energy dependent effects mentioned above are taken into account automatically: the synchrotron cooling of all secondary
 species is included, as it is important for the accurate prediction of the flavor ratios at the source.

The observed neutrino flavor composition at the detector is in general different from that at source
 due to neutrino flavor oscillations. Standard flavor mixing leads to achromatic (energy independent) transition
  probability for astrophysical neutrinos because of decoherence of the oscillations. For a pion beam source and standard flavor mixings, one
   gets an interesting prediction of flavor equipartition $1:1:1$ at the detector~\footnote{As long as
    $\theta_{23} \simeq \pi/4$ and $\theta_{13} \simeq 0$, this result  is robust and independent of
      the value of $\theta_{12}$.}.
It should be noted that there is no physical basis for this equipartition, it is purely accidental and just a
consequence of the specific choice of source and the mixing matrix in the near tri-bimaximal form along with the
assumption of standard mass-induced oscillations. Hence if we consider a source other than pion beam source
and/or non-standard propagation effects en-route, this prediction of universal ratio
changes~\cite{Pakvasa:2008nx}. This clearly implies that in order to infer new physics one has to be careful, and the mere departure from flavor equipartition can not serve as a guaranteed indicator of the specific class of new
physics in the most general situation.  The new physics effects studied in the context of high energy
astrophysical neutrinos include neutrino
decay~\cite{Beacom:2002vi,Majumdar:2007mp,Maltoni:2008jr,Bhattacharya:2009tx,Bhattacharya:2010xj}, pseudo-Dirac
nature of neutrinos~\cite{Beacom:2003zg,Bhattacharya:2010xj}, violation of discrete symmetries such as
CPT~\cite{Hooper:2004xr,Bhattacharya:2010xj,Bustamante:2010nq}, Lorentz invariance
violation~\cite{Hooper:2005jp,Bhattacharya:2009tx,Bhattacharya:2010xj}, quantum
decoherence~\cite{Hooper:2004xr,Morgan:2004vv,Hooper:2005jp,PhysRevD.72.065019,Bhattacharya:2010xj}, violation
of unitarity of the mixing matrix~\cite{Xing2008166}, coupling of neutrinos to dark energy~\cite{Ando:2009ts}
and  non-standard interactions~\cite{Blennow:2009rp}. However in the existing studies, the  energy dependence of
the flavor composition at the source has not been taken into account explicitly and many of these studies apply
to only a specific source type. In the present work, we incorporate
\begin{enumerate}
\item Energy dependent effects in the flavor composition at the source by using the HMWY model as our toy model, and
\item Energy dependent new physics scenarios during propagation of neutrinos from source to
detector.
\end{enumerate}
We study their impact on the detected flavor ratios for a point source. Most importantly, unlike the case of
standard oscillations which are achromatic, the energy dependent new physics effects lead to energy dependent
probabilities. Thus energy dependent observable flavor ratio is therefore an outcome of an interplay of two
energy dependent effects and one has to be careful to interpret these results for the new physics searches. In
order to elucidate this interplay, we explore two representative examples (a) neutrino decay,
 and (b) quantum decoherence.

The first non-standard possibility considered here is that of neutrino decay over astrophysical $L/E$ scales,
which sheds light on the lifetime of the neutrinos. Among the terrestrial experiments, solar neutrino data sets the strongest limit on neutrino lifetime to mass ratio $\tau^0/m \gtrsim 10^{-4}~ {\mathrm{s/eV}}$~\cite{Beacom:2002vi} which
implies that decay of high energy astrophysical neutrinos   can not be ruled out. Astrophysical neutrinos also
provide a much higher (by several orders of magnitude) neutrino lifetime sensitivity due to the higher $L/E$
ratio involved. Decay of astrophysical neutrinos has been studied  in
\Ref~\cite{Beacom:2002vi,Majumdar:2007mp,Maltoni:2008jr}. The other new physics scenario considered in the
present work is that of quantum decoherence effects. Even though to date there is no convincing theory of
quantum gravity, it is expected to give rise to distinctive signatures such as violation of Lorentz invariance,
CPT violation and/or quantum decoherence~\cite{Mavromatos:2007hv} at low energies (compared to the Planck
scale). Neutrino telescopes are particularly well-suited to probe such new physics effects in both
atmospheric~\cite{PhysRevD.79.102005} as well as astrophysical~\cite{Hooper:2005jp,Bustamante:2010nq} neutrino
fluxes. Note that some of these exotic effects  are related due to the CPT theorem, for example, CPT violation
implies Lorentz
       invariance violation but not vice-versa, and quantum decoherence can give rise to
        CPT violation~\cite{Mavromatos:2007hv}.
          In particular, quantum decoherence
         leads to evolution of pure states to mixed states via interaction with the environment of the
          space-time. The effect
           of quantum
     decoherence on astrophysical neutrinos has been studied in \Refs~\cite{Hooper:2004xr,Hooper:2005jp,PhysRevD.72.065019}.

The paper is organised as follows. In \Sec~\ref{sec:framework}, we introduce our theoretical flavor ratio
framework. Then in \Sec~\ref{sec:flr}, we briefly recapitulate the HMWY model and illustrate how the flavor
ratios of the sources acquire their energy dependence. We introduce our energy dependent new physics effects in
\Sec~\ref{sec:np}, where we also discuss their interplay with the energy dependent flavor ratios at the source
and detector. Finally, we perform a systematic scan of regions of the parameter space in \Sec~\ref{sec:hillas},
to identify
 source classes which may be most useful for this application. We then summarize in \Sec~\ref{sec:summary}.
Note that a more detailed discussion of the quantum decoherence model can be found in \App~\ref{app:decoh}.

\section{Flavor ratio framework}
\label{sec:framework}

We begin by defining the quantities of interest for the present study: the flavor ratio at the source, propagation effects, and the flavor ratio at the detector.

\subsection{Source flavor ratio} \label{sec:sframework}   Assuming a negligible amount
of $\nu_\tau$ at the source~\footnote{One may get a $\nu_\tau$ component in some exotic scenarios, such as cosmic
defects or evaporating black holes, typically at energies beyond $\sim 10^{19}$~eV. However, the HMWY model
 does not extend to such high energies in most cases because of the Hillas criterion~\cite{Hillas:1985is} and
  synchrotron cooling of the protons limiting the maximal energy. In addition, one typically expects a diffuse
   flux from such objects, where as for point sources, the angular resolution of the instrument can be used to suppress the backgrounds. Therefore, the contribution from these objects can be neglected if point source fluxes are considered.}, the flavor composition is completely
characterized by the ratio of the electron to muon neutrino flux
\begin{equation}
\xhat (E) = \frac{\Phi^{0}_e (E)}{\Phi^{0}_\mu (E)} \, , \label{equ:comp}
\end{equation} where $\Phi^{0}_e (E)$ and $\Phi^{0}_\mu (E)$ are the fluxes of electron and muon neutrinos
  without propagation effects. These correspond to the fluxes
  directly at the source, apart from an overall normalization, distance dependence, a possible Lorentz boost,
   and redshift effects. Since all these effects affect the different flavors in the same way, it
    is convenient to normalize to the fluxes at the detector without propagation effects
      $\Phi^{0}_\alpha (E)$. The fluxes are typically given in
units of $\mathrm{GeV^{-1} \, cm^{-2} \, s^{-1}}$ (point source) or
$\mathrm{GeV^{-1} \, cm^{-2} \, s^{-1} \, sr^{-1}}$ (diffuse flux). Note that we always refer to the sum
 of neutrino and antineutrino fluxes in the following, since we assume that the detector cannot
  distinguish between these.

In the literature, the following source classes are  distinguished depending upon the value of $\xhat$:
\begin{description}
\item[Pion beam sources] produce neutrinos from charged pion and successive muon decays, such as
\begin{eqnarray}
\pi^+ & \rightarrow & \mu^+ + \nu_\mu
 \, , \label{equ:piplusdec} \\
& & \hookrightarrow e^+ + \nu_e  + \bar\nu_\mu   \label{equ:muplusdec} \, .
\end{eqnarray}
This leads to  $\xhat \simeq 1/2$.
\item[Muon damped sources] produce neutrinos from pion decays only, \ie, \equ{piplusdec}, since the muons loose energy by synchrotron radiation efficiently. This leads to $\xhat \simeq 0$.
\item[Muon beam sources] The muons may pile up at lower energies, where muon decays dominate,
see \equ{muplusdec}. Then we have $\xhat \simeq 1$. Note that the neutrinos from semi-leptonic decays of charmed mesons in baryon rich astrophysical environments such as slow jet supernovae~\cite{Razzaque:2004yv} (where $pp$ interactions are prevalent) also lead to high energy neutrinos with $\xhat \simeq
1$~\cite{PhysRevD.79.053006}.
\item[Neutron beam sources] The neutrons, produced by photodissociation of heavy nuclei or photohadronic
interactions, decay into neutrinos by
\begin{equation}
 n \rightarrow  p + e^- + \bar{\nu}_e  \label{equ:ndec}  \, ,
\end{equation}
leading to $\xhat \gg 1$.
\item[Undefined sources]
 Several processes compete with similar magnitudes, leading to arbitrary $\xhat$.
\end{description}
Note that  $\xhat (E)$ is, in general, an energy dependent quantity, and these classifications only apply to
certain energy ranges. A given flavor composition at source uniquely identifies a particular mechanism provided
we have some information on the type of source giving rise to the neutrino flux. For instance, a muon beam
source can be mimicked by a transparent source (via decay of piled up muons at low energies)  or a hidden source
such as slow jet supernovae (via decay of charmed mesons) where different mechanisms are responsible for
arriving at $\xhat \simeq 1$. In addition, the guaranteed flux of extremely high energy neutrinos from
Greisen-Zatseptin-Kuzmin (GZK) process (photohadronic interaction of highest energy cosmic rays with cosmic
microwave background radiation) also yields energy dependent source flavor
ratios~\cite{Hooper:2004jc,Ave:2004uj}. Below 100 PeV, GZK neutrinos fall in neutron beam source class while
above 100 PeV, GZK neutrinos are more like a pion beam source.

\subsection{Standard propagation effects} \label{sec:sdframework}
 We assume propagation of astrophysical neutrinos in vacuum over distances long
enough such that oscillation effects become decoherent. On incorporating the propagation effects, the flavor
flux changes to $\Phi^{\mathrm{Det}}_\beta$ for neutrino of flavor $\nu_\beta$
\begin{equation}
\Phi^{\mathrm{Det}}_\beta(E) = \sum\limits_{\alpha=e,\mu,\tau}^{} P_{\alpha \beta}(E) \, \Phi^{0}_\alpha(E) \, ,
\label{equ:flmix}
\end{equation}
where for standard oscillations, we have \begin{equation}
  P_{\alpha \beta}(E) \equiv P_{\alpha \beta} = \sum\limits_{i=1}^{3}  |U_{\beta i}|^2 \, |U_{\alpha i}|^2 \,
\, , \label{equ:pso}
\end{equation}
which is achromatic. Note that \equ{pso} depends only on the mixing angles $\theta_{ij}$ contained in the mixing
matrix elements $U_{\beta i}$  of the $3 \times 3$  Pontecorvo-Maki-Nakagawa-Sakata (PMNS) neutrino mixing
matrix which in the commonly adopted Particle Data Group (PDG) parametrization is given by
\begin{equation}
U = \begin{pmatrix} c_{12}c_{13} & s_{12} c_{13} & s_{13} e^{-i \delta_{CP}} \\-s_{12} c_{23} -c_{12} s_{23}
s_{13} e^{i \delta_{CP}}& c_{12} c_{23} -s_{12} s_{23} s_{13} e^{i \delta_{CP}}& s_{23} c_{13} \\ s_{12} s_{23}
- c_{12} c_{23} s_{13}e^{i \delta_{CP}} & -c_{12} s_{23} -s_{12} c_{23} s_{13} e^{i \delta_{CP}}& c_{23} c_{13}
\end{pmatrix}\, ,
\end{equation}
where $s_{ij}=\sin \theta_{ij}$ and $c_{ij}=\cos \theta_{ij}$ and $\delta_{CP}$ is the Dirac type CP phase.
Unless noted otherwise, we use the following values for the mixing parameters, see, \eg, \Ref~\cite{Schwetz:2008er}:
\begin{equation} \sin^2 \theta_{23}=0.5, \quad \sin^2 \theta_{12}=0.318 , \quad  \sin^2
\theta_{13}=0\,.\label{equ:bf}
\end{equation}
  Thus, we
 note that  flavor equipartition  at the detector $\Phi_{e}^{\mathrm{Det}}:\Phi_{\mu}^{\mathrm{Det}}:
    \Phi_{\tau}^{\mathrm{Det}}=1:1:1$ is a consequence of two inputs:
     the source ratio $\xhat \sim 0.5$ (pion beam source) and the use of mixing angles being close to the
       tri-bimaximal form.

\subsection{Flavor ratio at  the detector} \label{sec:frframework}
We need to take into account the propagation effects, before computing the observable flavor ratios at the
detector.  Ideally one would want to detect all the flavors separately, however in practice it is not so easy.
Several flavor ratios have been constructed in literature to distinguish between flavors~\cite{Pakvasa:2010jj}.
The easiest possibility to measure flavor ratios at neutrino telescopes requires, apart from muon tracks
sensitive to $\nu_\mu$,  the identification of cascades~\cite{IcCascade:2011ui}. These come with a lower statistics and have a higher threshold (about 1 to 10~TeV in IceCube). In addition, the neutrino effective area increases much weaker with energy, which means that the best statistics may be obtained close to the threshold. However, one expects a much better energy resolution for cascades.  If we assume that electromagnetic (from $\nu_e$) and hadronic (from $\nu_\tau$)
cascades do not need to be distinguished, a useful observable is the ratio of muon tracks to
cascades~\cite{Serpico:2005sz}
\begin{eqnarray}
\rhat &\equiv& \frac{\Phi_{ \mu}^{\mathrm{Det}}}{\Phi_{e}^{\mathrm{Det}}+\Phi_{\tau}^{\mathrm{Det}}}
= \frac{P_{e\mu}(E)  \, \xhat (E) + P_{\mu\mu}(E)  }{[P_{ee}(E)+P_{ e \tau}(E) ]\, \xhat (E) + [P_{\mu
e}(E)+P_{\mu \tau}(E)]} \, .
 \label{equ:R}
\end{eqnarray}
Note that the above formula holds even if unitarity is violated, \ie, $P_{e \alpha} + P_{\mu \alpha} + P_{\tau
\alpha} < 1$, such as for neutrino decay into invisible states.
 In addition, note that neutral current events will also produce cascades, which, in practice, have
to be included as background. In \Ref~\cite{IcCascade:2011ui}, the most recent IceCube cascade analysis, the contribution of the different flavors for a $E^{-2}$ extragalactic test flux with equal contributions of all flavors at the Earth was given as: electron neutrinos 40\%, tau neutrinos 45\%, and muon neutrinos 15\% (after all cuts). This implies that charged current showers dominate and that electron and tau neutrinos are detected with comparable efficiencies, \ie,  that \equ{R} is a good first approximation to discuss flavor at a neutrino telescope.
The benefit of this flavor ratio is that the
  normalization of the source drops out. In addition, it represents the experimental flavor measurement with the simplest possible assumptions.
    \equ{R} implies that the observable  flavor flux ratio $\rhat$
  depends on two energy dependent quantities: $\xhat(E)$ characterizing the energy dependence of the flavor
composition at the source, and $P_{\alpha \beta} (E)$ characterizing the energy dependence during the
propagation from source to detector. In the absence of energy dependent new physics, the probability is
independent of energy (see \equ{pso}) and the flavor ratio $\rhat$ closely follows the energy dependence of the
source
 function $\xhat (E)$.

Using \equ{pso} in \equ{R}, we have listed the characteristic values of flavor ratios $\xhat,\rhat$  in
\Tab~\ref{tab:sourceclass} assuming standard oscillations and best-fit values of oscillation parameters.  In
this study, we do not consider the uncertainties coming from the oscillation parameters, since we expect that
the discussed measurements are only feasible with sufficient statistics on a timescale when the oscillation parameters are sufficiently
limited; see discussion in \Ref~\cite{Hummer:2010ai} (Fig.~11).

\begin{table}[]
\begin{center}
\begin{tabular}{l  r   c }
\hline
\\[-0.45cm]
\hspace{0.2cm}{Source}\hspace{0.2cm} & \hspace{0.2cm}{${\xhat}$}\hspace{0.2cm} & \hspace{0.2cm}
{${\rhat}$}\hspace{0.2cm}
\\
\\[-0.45cm]
        \hline
        Muon damped  & $0$ & 0.64
          \\
                 Pion beam & $0.5$   &   0.50 \\
        Muon beam  & $1$ & 0.44
          \\
        Neutron beam & $\gg 1$ & 0.28
         \\
        Undefined & Other & Other \\
          \hline
\end{tabular}
\mycaption{ \label{tab:sourceclass} Characteristic values of flavor ratios $\xhat$ and $\rhat$ for the
considered source classes for best-fit values of oscillation parameters. The ratio $\rhat$ is computed using
\equ{R} and standard oscillation probabilities (\equ{pso}).  }
\end{center}
\end{table}

\section{Energy dependent flavor ratios at the source}
\label{sec:flr}

The HMWY model describes neutrino production via photohadronic ($p\gamma$) processes for transparent sources
(optically thin to neutrons) and includes the cooling of the secondary particles. It can be used to generate
neutrino fluxes as a function of few astrophysical parameters. Below we outline the key ingredients of the
model and the main results, for details  see \Ref~\cite{Hummer:2010vx,Hummer:2010ai}. All of the following
quantities refer to the frame where the target photon field is isotropic, such as the shock rest frame (SRF).

\subsection{Ingredients of the HMWY model}
The protons and electrons/positrons are injected with spectra $\propto E^{-\alpha}$. The maximal  energies of
these spectra are determined by balancing the energy loss and acceleration timescale given by
  \begin{equation}
  t^{-1}_\mathrm{acc}=\eta\frac{c^2 e B}{E} \,,
 \end{equation}
with $\eta$ an acceleration efficiency depending on the acceleration mechanism, where we choose $\eta=0.1$
later. If synchrotron losses dominate, the maximal energy is therefore given by
\begin{equation}
E_{\mathrm{max}} = \sqrt{ \frac{9 \pi \epsilon_0 m^4 c^7 \eta}{e^3 B}} \, .
\end{equation}
It scales $\propto m^2$, which means that the protons are accelerated to much higher energies, and $\propto 1/\sqrt{B}$, which means that strong magnetic fields limit the maximal energies.
For each particle species, the injection and energy losses/escape are balanced by the steady state equation
\begin{equation}
\label{equ:steadstate}
Q(E)=\frac{\partial}{\partial E}\left(b(E) \, N(E)\right)+\frac{N(E)}{t_\mathrm{esc}} \, ,
\end{equation}
with $t_\mathrm{esc}(E)$ the characteristic escape time, $b(E)=-E\,t_\mathrm{loss}^{-1}$ with
$t_\mathrm{loss}^{-1}(E)=-1/E \, dE/dt$ the rate characterizing energy losses, $Q(E)$ the particle injection
rate $[\mathrm{\left( GeV\,s\,cm^3\right)^{-1}}]$ and $N(E)$ the steady particle spectrum $[\mathrm{\left(
GeV\,cm^3\right)^{-1}}]$. For all charged particles, synchrotron energy losses  and adiabatic cooling  are
taken into account.  In addition, unstable secondaries, \ie, pions, muons, and kaons, may escape via decay. As a
consequence, for pions, muons, and kaons, neglecting the adiabatic cooling, the (steady state) spectrum is
loss-steepend above the energy
\begin{equation}
E_c = \sqrt{ \frac{9 \pi \epsilon_0 m^5 c^5}{\tau_0 e^4 B^2}} \, ,
\label{equ:ec}
\end{equation}
where synchrotron cooling and decay rates are equal. One can read off this formula that the different
secondaries, which have different masses $m$ and rest frame lifetimes $\tau_0$,  will exhibit different break
energies $E_c \propto \sqrt{m^5/\tau_0}$ which solely depend on particle physics properties and the value of
$B$.

\begin{figure}[t]
 \includegraphics[width=1\textwidth,viewport=20 192 696 548]{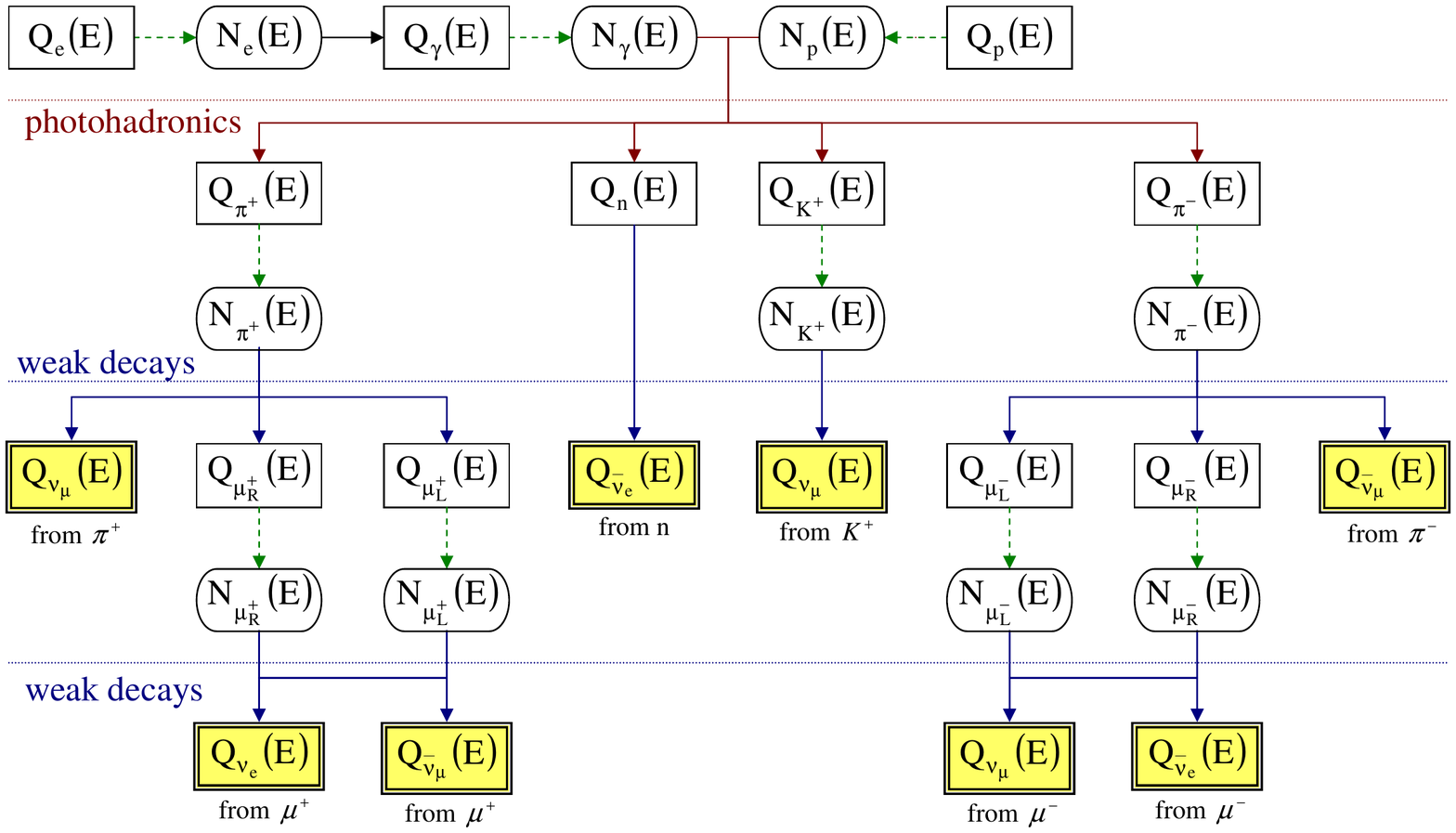}
 \mycaption{\label{fig:flowchart}Flowchart describing the neutrino production in the HMWY model. The functions $Q(E)$ denote (injection) spectra per time frame $[\mathrm{\left( GeV\,s\,cm^3\right)^{-1}}]$ and $N(E)$ steady spectra $[\mathrm{\left( GeV\,cm^3\right)^{-1}}]$ derived from the balance between injection and losses or escape. Dashed arrows stand for solving the steady state differential equation \equ{steadstate}, the horizontal line ``photohadronics'' to solving \equ{prodmaster} for all interaction types.
Figure taken from \Ref~\cite{Hummer:2010ai}.}
\end{figure}

While being accelerated, the electrons loose energy into synchrotron photons, which serve as the target
photon field. Charged meson production then occurs via
\begin{align}
 p + \gamma & \rightarrow \pi + p' \, ,\\
 p + \gamma & \rightarrow K^+ + \Lambda/\Sigma \, ,
\end{align}
with these synchrotron photons, where the leading kaon production mode is included and $p'$ is a  proton or
neutron.  In addition, two- and multi-pion production processes are included (not listed here), see \Ref~\cite{Hummer:2010vx}
for details. The injection of the charged mesons is computed from the steady state proton $N_p(E_p)$ and photon
$N_\gamma(\varepsilon)$ spectra with~\cite{Hummer:2010vx}
\begin{equation}
Q_b(E_b) = \int\limits_{E_b}^{\infty} \frac{dE_p}{E_p} \, N_p(E_p) \, \int\limits_{\frac{\epsilon_{\mathrm{th}}
  m_p}{2 E_p}}^{\infty} d\varepsilon \, N_\gamma(\varepsilon) \,  R_b( x,y ) \,,
\label{equ:prodmaster}
\end{equation}
with $x=E_b/E_p$ the fraction of energy going into the secondary, $y \equiv (E_p\varepsilon)/m_p$  (directly
related to the center of mass energy), a ``response function'' $R_b( x,y )$ (see \Ref~\cite{Hummer:2010vx}), and
$\epsilon_{\mathrm{th}}$ the threshold for the photohadronic interactions (in terms of photon energy in the   proton rest
frame). The weak decays of the secondary mesons, such as Eqs.~(\ref{equ:piplusdec}) and (\ref{equ:muplusdec}),
are described in \Ref~\cite{Lipari:2007su}. These will finally lead to neutrino fluxes from pion, muon, kaon,
and neutron decays. The whole procedure is illustrated in \figu{flowchart}, the finally obtained eight neutrino  fluxes are shown as shaded boxes. In the following, we will sum over the fluxes from all polarities. For instance, if we refer to ``$\nu_\mu$ from muon decays'', we mean the sum over the $\nu_\mu$ from $\mu^-$ and the $\bar\nu_\mu$ from $\mu^+$ in \figu{flowchart}.

One can clearly see from \equ{prodmaster}, that only the product of the proton and photon  (and therefore
electron) density normalizations enters the final result, as long as cooling processes implicitly depending on
proton-electron ratio, such as inverse Compton scattering or photohadronic cooling, are sub-dominant. This is,
of course, a key simplifying assumption which, in general, limits the applicability of this model. However, it
leads to the simplest possible model which includes the effects relevant for the flavor ratios, whereas
additional cooling processes affecting the protons and electrons only change the shape of the neutrino spectra
(including maximal energies), but not the flavor ratios as a function of energy directly. In the flavor ratios,
the  overall normalization from \equ{prodmaster} cancels. In that sense, the flavor composition might be the
most robust prediction one can make for a source, since it is very little affected by astrophysical
uncertainties as long as the dominating cooling process of the secondaries (pions, muons, kaons) is synchrotron
emission, and the dominant escape process is decay. These processes depend, however, on the magnetic field in
the source $B$, a quantity, which is not directly accessible.

In the following, for the sake of simplicity, we assume that we can estimate the parameters of the source by
utilizing the multi-messenger connection (such as gamma-ray observations). For example, one can estimate the
magnetic field $B$ from energy equipartition, the injection index from the spectral shape, and the size of the
acceleration region $R$ from the variability timescale. Then one can predict the flavor ratios as a function of
these parameters. Conversely, flavor ratio measurements may provide a direct handle on astrophysical quantities
such as the magnetic field~\cite{Hummer:2010ai}. We use $\alpha=2$, since we do not obtain strongly qualitatively
different results in the range $2 \lesssim \alpha \lesssim 3$.

\subsection{Hillas criterion and source classification on the Hillas plot}

\begin{figure}[tp]
\centering
\includegraphics[width=\textwidth]{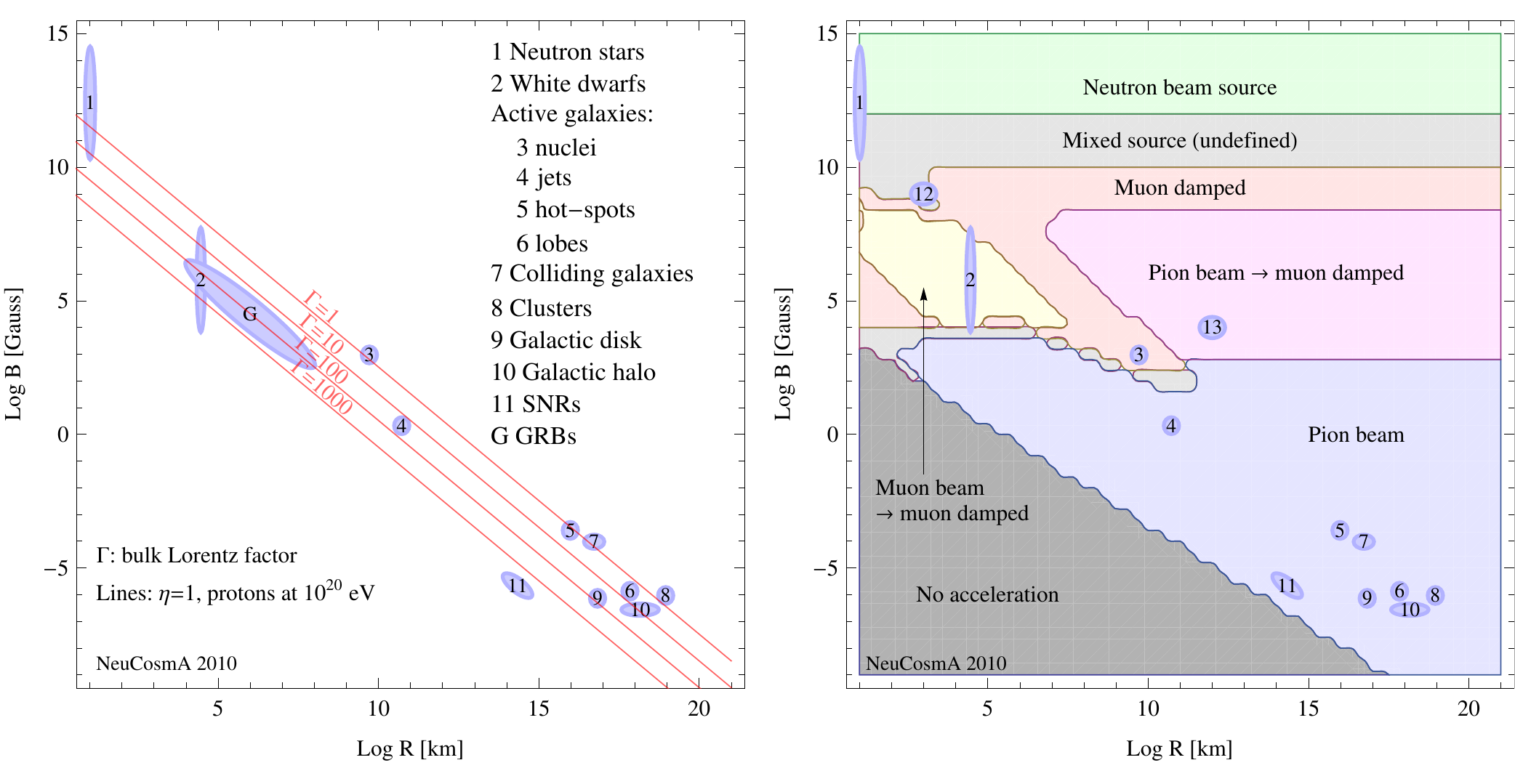}
\mycaption{\label{fig:hillas} Left panel: Possible acceleration sites in Hillas plot as a function of $R$ and $B$ (version adopted from M. Boratav). Right panel: Classification of sources for injection index $\alpha=2$ in this plot (see main text). Some points from left plot are shown for orientation, as well as two new points (12 and 13) are defined for later reference. Figure taken from \Ref~\cite{Hummer:2010ai}.}
\end{figure}

The parameters $R$ and $B$ can be directly related to the Hillas plot, see \figu{hillas}, left panel, for  an
example. In order to confine a particle in a magnetic field at the source, the Larmor radius has to be smaller
than the extension of the acceleration region $R$. This can be translated into the Hillas condition for the
maximal attainable energy~\cite{Hillas:1985is}
\begin{equation}
E_{\mathrm{max}} \, [\mathrm{GeV}] \simeq 0.03 \cdot \eta \cdot Z \cdot  R \, [\mathrm{km}] \cdot B
 \, [\mathrm{G}] \, ,
\label{equ:hillas}
\end{equation}
where $Z$ is the charge (number of unit charges) of the accelerated particle, $B$ is the magnetic field  in
Gauss, and $\eta$ can be interpreted as an efficiency factor or linked to the characteristic velocity of the
scattering centers. One complication in this type of figure is that $R$ and $B$ in \equ{hillas} are potentially
given in the SRF, whereas $E_{\mathrm{max}}$ is to be determined in the observer's frame. If the particles are
accelerated in a relativistically moving environment, such as in a GRB fireball, this assignment is not trivial
anymore, and the relatively large Lorentz boost $\Gamma$ of the acceleration region into the observer's frame must be taken into account. We stick to the
interpretation of $R$ and $B$ in the SRF, which means that (for $\eta=1$) the condition in \equ{hillas} depends
on the Lorentz boost of the source. This is illustrated by showing \equ{hillas} for several selected Lorentz
boosts in \figu{hillas}, left panel, for protons, $\eta=1$, $E_{\mathrm{max}}=10^{20} \, \mathrm{eV}$. In the
following, we will use $\Gamma=1$. However, it should be noted that if the source is significantly boosted, the neutrino energies will have to be increased by $\Gamma$, and the (cascade) threshold of the neutrino telescope may be passed. We neglect redshift effects on the other hand.

In \figu{hillas}, right panel, the main result of \Ref~\cite{Hummer:2010ai} is shown for $\alpha=2$: Here the
sources are classified into the categories from Sec.~\ref{sec:framework} as a function of $R$ and $B$.
 Note that a given source may fall into different categories in different energy ranges.
  For example, the ``muon beam $\rightarrow$ muon damped'' region means that a muon beam is found for low
   energies, whereas a muon damped source is found for higher energies. The classical ``pion beam
    $\rightarrow$ muon damped'' category is found for large enough magnetic fields and large
     enough $R$ as well, where the transition between adiabatic and synchrotron cooling leads to an additional break in the proton spectrum.
\begin{figure}[tp]
\centering
\includegraphics[width=\textwidth]{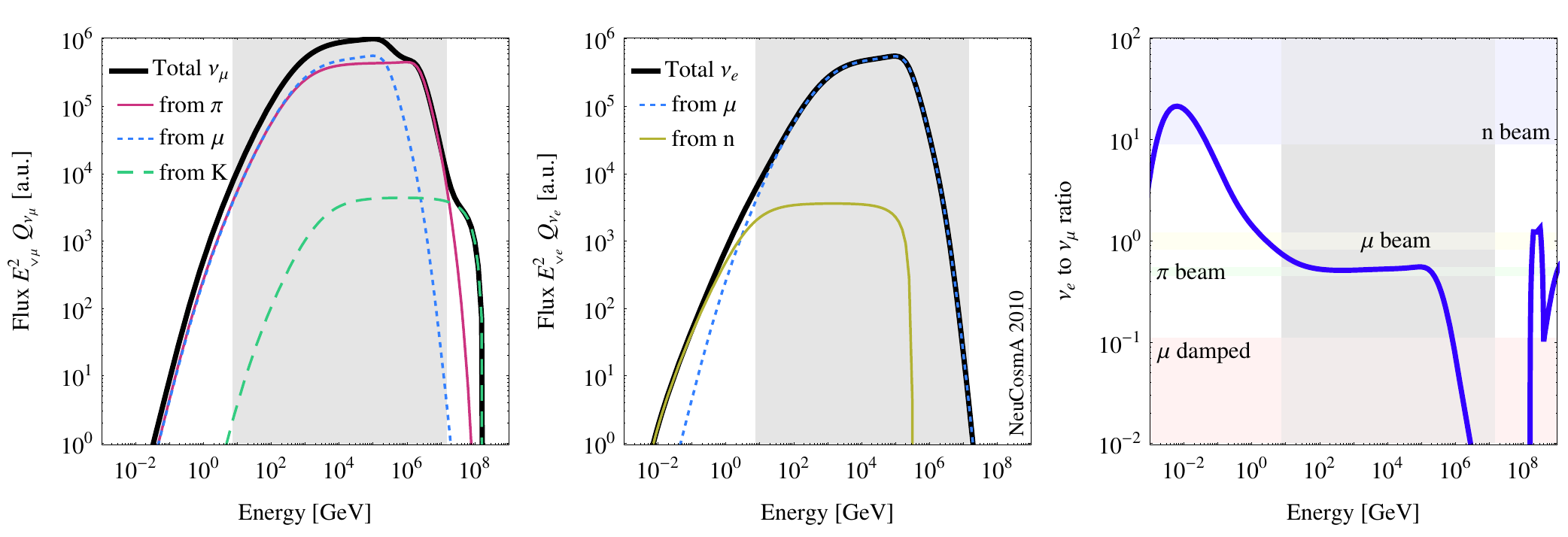}
\mycaption{\label{fig:tp13} Muon (left panel) and electron (middle panel) neutrino fluxes at the source, and $\xhat$ (right
panel) as a function of energy for test point (TP)~13 from \figu{hillas}, right panel. The contributions of the individual
neutrino fluxes from different parents are shown as well. The gray-shaded regions mark the energy window used for the source classification (see main text). Figure adopted from \Ref~\cite{Hummer:2010ai}.}
\end{figure}

We illustrate the classification of the sources for test point (TP)~13 from \figu{hillas} (right panel), see \figu{tp13},  where the neutrino fluxes and flavor ratio at the source are explicitly shown for an injection index, $\alpha=2$.
One can easily see in the left panel that the different cooling
and decay timescales of the secondaries lead to a hierarchy of the breaks, as described in \equ{ec}, an effect
which is similarly present in GRBs (see, \eg, \Refs~\cite{Lipari:2007su,Baerwald:2010fk}). The $\nu_e$ flux
mostly comes from muon decays (middle panel), which means that it will roughly follow the muon break as in the
left panel, whereas $\nu_\mu$ from pion and kaon decay (left panel) extend to higher energies. As a consequence,
the flavor ratio $\xhat$ changes from $0.5$ to $0.0$ in the right panel at about $10^6 \, \mathrm{GeV}$. The
neutrinos from neutrons, which are produced in the photohadronic interactions, lead to a $\nu_e$ flux dominating
at low energies. However, in that range the total flux is already too low compared to the peak and also drowned
in the atmospheric neutrino background.

Of course the above identification of different source types  applies only  to  an energy range where the flux
of at least one of the flavors is large enough, to give reasonable statistics.  Since the region close to peak
in $E^2 \Phi_\alpha^{0}$ will contribute mostly to the event rates, at least in energy ranges where the neutrino
effective area is proportional to $E^2$, we define an energy window which captures the ``upper two orders of
magnitude" in the flux. That is, we compute (for each point) the maximal flux in $E^2 \Phi_\alpha^{0}$, and
derive the energy range where the flux of any flavor is at least $1\%$ of the respective maximal flux (in  $E^2
\Phi_\alpha^{0}$). Only this energy range is considered, since otherwise the flavor ratio may be ill-defined.
This energy range is marked as shaded regions in \figu{tp13}. In the right panel, one can also see that the
flavor ratio is ill-defined above $10^8 \, \mathrm{GeV}$, where the fluxes are negligible. In addition, note
that it is required that a specific flavor ratio category is found over at least one order of magnitude in
energy within the chosen energy window.

\section{Energy dependent new physics scenarios}
\label{sec:np}

In presence of new physics, the flavor ratio $\rhat$ not only depends on energy dependent flavor composition of
the source $\xhat (E)$, but also on the new physics induced energy dependent terms in the probability,
see~\equ{flmix}. We consider two specific examples of new physics here which can give rise of energy dependent
effects in opposite energy regimes. Whereas neutrino decays are mostly present at lower energies (the lifetime
is Lorentz-boosted), quantum decoherence may plausibly cause the strongest effects at
  higher energies. The energy dependence of source is contained in $\xhat(E)$ as described in
   Sec.~\ref{sec:flr} while the energy dependent new physics effects lead modification of
     probabilities relevant for astrophysical neutrinos  as described below.

\subsection{Neutrino decay} Neutrino decay is usually described by an energy dependent
damping of the overall oscillation probability~\cite{Beacom:2002vi} and \equ{pso} gets modified to
\begin{equation}
 P_{\alpha \beta} = \sum\limits_{i=1}^{3}  |U_{\beta i}|^2 \, |U_{\alpha i}|^2 \,
 \, D_i(E) \,  \quad \text{with} \quad D_i(E) =
\exp \left( - \hat\alpha_i \frac{L}{E} \right)\,, \label{equ:pdecay}
\end{equation} as
the {\em damping coefficient}~\footnote{In the spirit of \Ref~\cite{Blennow:2005yk}, see Eqs.~(2.3)
 and~(2.4),
which contain different damping effects. For astrophysical neutrinos, we can only probe $D_{ii} \equiv D_{i}$.}.
Here $\hat\alpha_i=m_i/\tau^0_i$ with $\tau^0_i$ is the rest frame lifetime for mass eigenstate $\nu_i$. Typically
the neutrino lifetime is quoted as $\tau_i^0/m_i$   since $m_i$ is unknown. From the exponential factor, the
neutrinos decay if
\begin{eqnarray}
\frac{\tau_i^0}{m_i} = \hat\alpha_{i}^{-1} & \lesssim & 10^2 \,\frac{L}{\mathrm{Mpc}} \frac{\mathrm{TeV}}{E}
~{\rm{s \, eV^{-1}}}\,.
\end{eqnarray}
  Thus, neutrino telescopes can probe lifetimes of the order of $\sim 10^{2} ~{\mathrm{s \,eV}}^{-1}$ for $L \sim 1$ Mpc ($\simeq
10^{14}$ s) and $E \sim 1$ TeV. In what follows, we use the same lifetime for all mass eigenstates,
$\tau_i^0/m_i \equiv \tau^0/m$ for the sake of simplicity.

In view of the rather weak direct limits on neutrino lifetime and the proposed unparticle scenarios, one can
have many different mechanisms of neutrino decay. In what follows, we do not assume a particular decay scenario
but consider the general case~\cite{Maltoni:2008jr}. For neutrino decay, one typically distinguishes:
\begin{description}
\item[Visible decays] One of the decay products is visible (to the neutrino detector), typically a lighter active neutrino
mass eigenstate.
\item[Invisible decays] The decay products are invisible (to the neutrino detector), such as sterile neutrinos or unparticles.
\end{description}
In addition, we have
\begin{description}
\item[Incomplete decays] The mass eigenstates have decayed partially, \ie, $0 \le D_i \le 1$.
\item[Complete decays] The mass eigenstates have decayed completely, \ie, $D_i \rightarrow 0$.
\end{description}
In our framework \equ{pdecay}, we can describe complete or incomplete invisible decays.  For a framework for
visible complete decays, see \Ref~\cite{Maltoni:2008jr}. For the most general framework, see
\Refs~\cite{Lindner:2001fx,Lindner:2001th}. Note that if only a single mass eigenstate is stable in complete
decays, $\rhat$ does not depend on $\xhat(E)$ since probabilities factorize in the source-dependent
 and detector-dependent parts; \cf, \Refs~\cite{Pakvasa:1981ci,Maltoni:2008jr}.
\begin{figure}[t!]
\begin{center}
\includegraphics[width=0.5\textwidth]{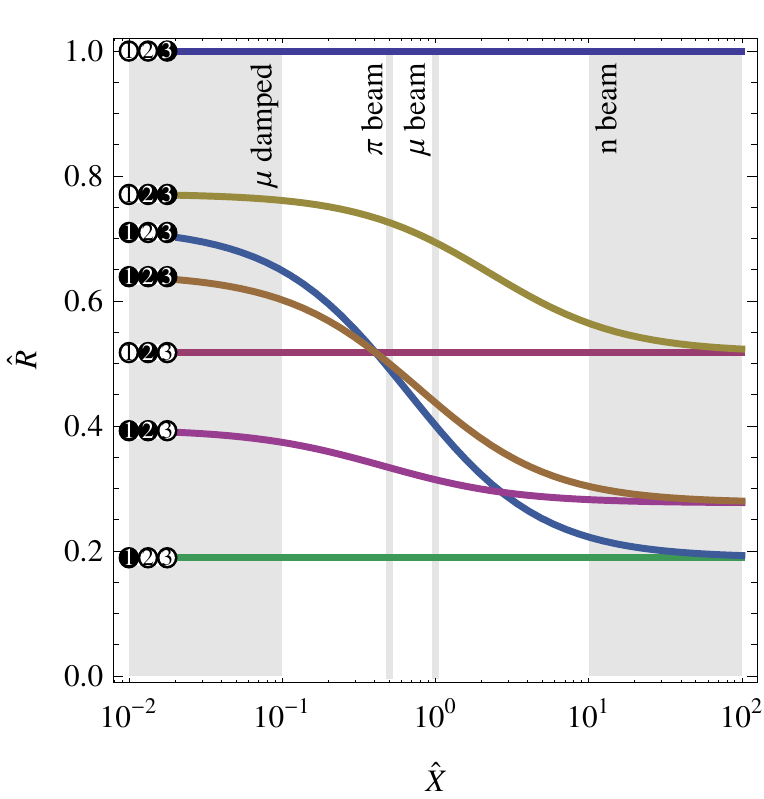}
\end{center}
\mycaption{\label{fig:xdecay} Flavor ratio $\rhat$ as a function of $\xhat$
for all complete decay scenarios. Black disks refer to stable mass eigenstates, white disks
to unstable mass eigenstates. Different sources classes as a function of $\xhat$
are marked.}
\end{figure}
In general, any decay scenario is characterized by stability properties of active states  which gives in all
$2^3=8$ scenarios for invisible decays, since each mass eigenstate can be either stable or not. We do not
consider the case of only unstable states, since no signal will be observable then.

 We show in \figu{xdecay}
the flavor ratio $\rhat$ as a function of $\xhat$ for all complete decay scenarios.  Black disks refer to stable
mass eigenstates, and white disks to  unstable mass eigenstates. One can read off this figure, that for the pion
beam, four scenarios can be clearly separated, whereas three scenarios are almost degenerate. If, however, the
information from a muon damped sources is added, such as at high energies, exactly these three scenarios, in
principle, separate. The muon beam has a similar effect in the opposite direction of $\xhat$. Therefore, it is
crucial to discuss sources which contain information from different parts of $\xhat$.

\begin{figure}[t!]
\begin{center}
\includegraphics[width=\textwidth]{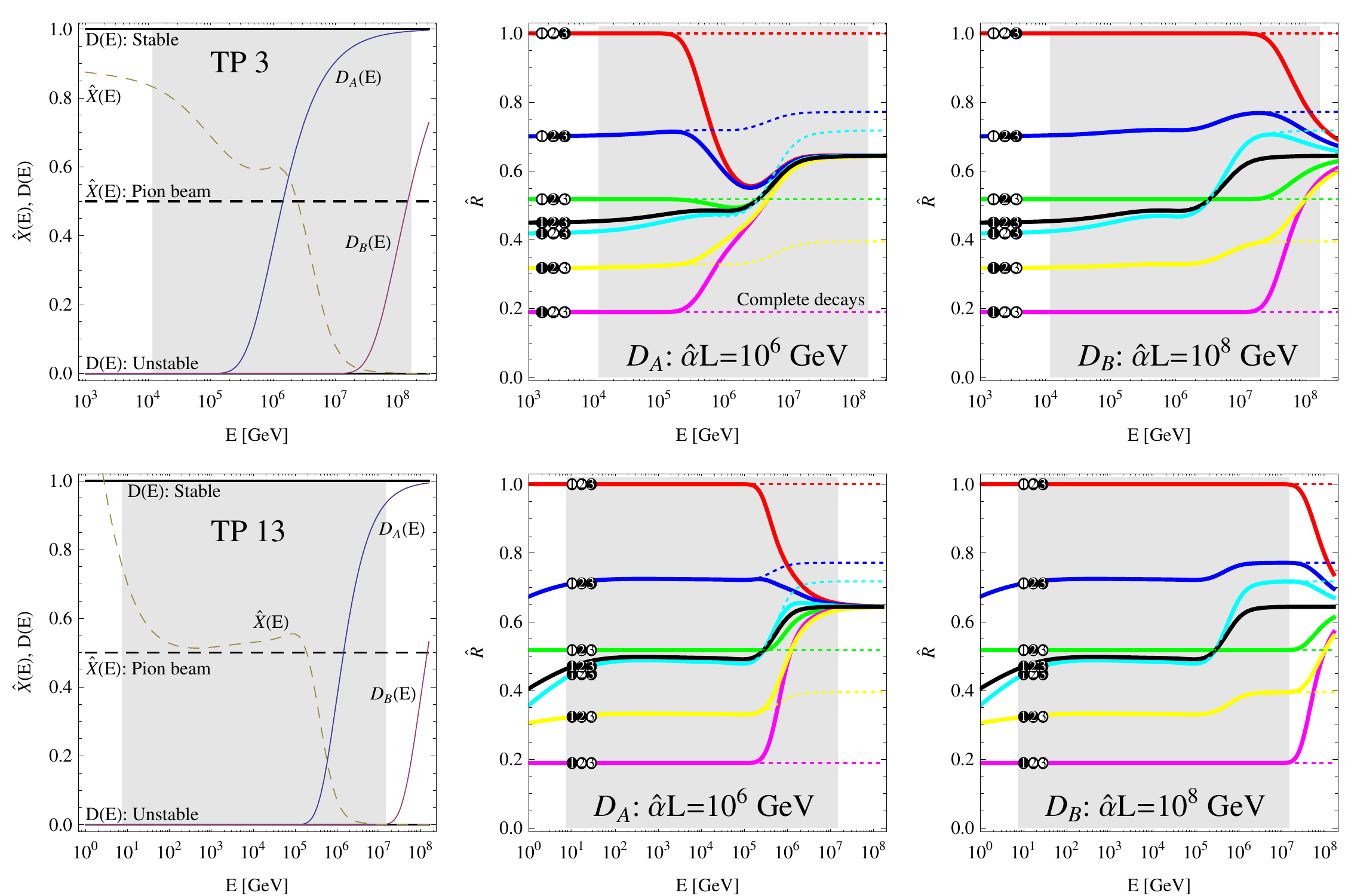}
\end{center}
\mycaption{\label{fig:decayall} Left panels: Energy dependence of the damping coefficients for $\hat\alpha L=10^6 \,
\mathrm{GeV}$ ($D_A$) and for  $\hat\alpha L=10^8 \, \mathrm{GeV}$ ($D_B$), as well as for the
 function $\xhat(E)$ (dashed curves).
The upper row corresponds to TP~3 from \figu{hillas}  (AGN nuclei), the lower row to TP~13. Middle and right
panels: Energy dependence (solid curves) of the flavor ratio $\hat R$  for seven different decay scenarios.
Black disks refer to stable mass eigenstates, white disks to unstable mass eigenstates. The middle panels are
plotted for $\hat\alpha L=10^6 \, \mathrm{GeV}$ ($D_A$) and the right panels for $\hat\alpha L=10^8 \, \mathrm{GeV}$
($D_B$), where the same decay rate is chosen for all unstable mass eigenstates. The region where the flux is
relatively large is shown as shaded region.   The dotted curves refer to complete decays. }
\end{figure}

We show in \figu{decayall} two  anomalous sources described in Sec.~\ref{sec:flr}: TP~3 (upper row) and TP~13
(lower row) from \figu{hillas}. In the left panels, the energy dependent damping coefficients $D_A$ (for $\hat\alpha
L=10^6 \, \mathrm{GeV}$) and $D_B$ (for $\hat\alpha L=10^8 \, \mathrm{GeV}$) are shown, as well as the flavor ratio
at the source $\xhat(E)$~\footnote{Note that the chosen decay rates correspond to lifetimes of about $0.1 \,
\mathrm{s \, eV^{-1}}$ ($D_A$) and $10^{-3} \, \mathrm{s \, eV^{-1}}$ ($D_B$) for $L=1 \, \mathrm{Mpc}$ using
$\tau^0/m \sim \hat\alpha ^{-1}$.
 These lifetimes are at least one order
of magnitude above the current direct bounds (depending on the mass eigenstate) except from the SN 1987A bound
(see, \eg, \Ref~\cite{Pakvasa:2010jj} and references therein). It is, however, strictly speaking not trivial to
assign the SN 1987A bound to $\nu_1$ or $\nu_2$, given the low statistics; the problem is that a flavor
eigenstate ($\bar \nu_e$) is measured which contains significant contributions from $\nu_1$ and $\nu_2$. The
only conclusion from that observation is that it is unlikely that both $\nu_1$ and $\nu_2$ are unstable at that
energy-distance scales, which means that the upper curves in \figu{decayall} (middle and right panels) are
basically excluded.}.
One can read off these panels, that decays become effective below $10^6 \, \mathrm{GeV}$ ($D_A$) or $10^8 \,
\mathrm{GeV}$ ($D_B$) for our choice of parameters. Well above these energies, the mass eigenstates are
practically stable (the lifetime increases with the Lorentz boost of the particles). The curves of $\xhat(E)$
clearly indicate that both the sources TP~3 and TP~13 (upper left and lower left panel) perform as muon-damped
sources at high energies for which $\xhat \simeq 0$. At lower energies, TP~3 is undefined, whereas TP~13
performs, within reasonable uncertainty, as pion beam.

In the middle and right panels of \figu{decayall}, the flavor ratio $\rhat $  is shown for seven different decay
scenarios (solid curves).  The middle panel corresponds to $D_A$ while the right column to $D_B$. As mentioned
before, the same decay rate is chosen for all unstable mass eigenstates. Note that $\xhat(E)$ and $\rhat(E)$
have opposite behavior as a function of $E$. Let us first focus on plots corresponding to $D_A$ (middle panel).
 From the left panel for TP~3 and TP~13, we can read off that the $D_A$ and $\xhat$ curves intersect at about the
same energy, which means that neutrino decays are only effective in the undefined (TP~3) or pion beam (TP~13)
energy range below about $10^6 \, \mathrm{GeV}$. From \figu{xdecay}, we know that this means that some of the
decay scenarios cannot be clearly separated, which we indeed find in the lower energy range of the middle
columns of \figu{decayall}. For $E \gg 10^6 \, \mathrm{GeV}$, however, all mass eigenstates are stable, which
can be seen from the deviation of the solid curves with respect to the dotted ones which refer to complete
decays (independent of energy). If all mass eigenstates are stable, the same physics scenario is recovered in
all cases, and no new information is added. This is different for the higher decay rate in the right column:
here the muon damped nature of the source allows for a separation of the different curves in energy range $E
\simeq 10^6-10^8 \, \mathrm{GeV}$, where both the test points evolve into muon-damped source type, but the mass
eigenstates are not yet stable. In this range, the degenerate curves separate out. Finally, \figu{decayall} is
an example for the interplay between the source characteristics $\xhat(E)$ and the decay-type new physics
scenario.

\subsection{Quantum decoherence}
\label{sec:qd}

The impact of quantum decoherence on the flux of high energy astrophysical neutrinos has been studied in  \Refs~\cite{Hooper:2004xr,Hooper:2005jp,PhysRevD.72.065019,Bhattacharya:2010xj}.
    We show in \App~\ref{app:decoh} how and under what conditions the oscillation
      probabilities presented for three neutrino flavors in this
     section are obtained. Under the
     assumption that oscillations average out over astrophysical distances, we can express the transition
      probability as a function of only two non-zero
decoherence parameters $\Psi$ and $\delta$,
\begin{eqnarray}
 P_{\alpha\beta} = \frac{1}{3} +
\frac{1}{2} (U_{\alpha 1}^2 - U_{\alpha 2}^2)(U_{\beta 1}^2-U_{\beta 2}^2)
 D_{\Psi} +
 \frac{1}{6}  (U_{\alpha 1}^2+U_{\alpha 2}^2- 2
 U_{\alpha 3}^2)(U_{\beta 1}^2+U_{\beta 2}^2 - 2 U_{\beta 3}^2)  D_{\delta}
 \, ,
 \label{equ:pdecoh}
\end{eqnarray}
where $D_{\Psi}$ and $D_{\delta}$ are the damping factors (corresponding to the eigenvalues of $\lambda_3$ and $\lambda_8$, respectively, of the
  decoherence matrix described in \App~\ref{app:decoh}) given by~\footnote{Note that, $\kappa$ is used to denote $\Psi,\delta$ and
   compared to  \App~\ref{app:decoh}, we incorporate the energy dependence explicitly here, \ie, we
    replace $\Psi \rightarrow \Psi E^n$, $\delta \rightarrow \delta E^n$.}
 \begin{equation}
   D_\kappa (E) =
\exp \left( - 2 \, \kappa \, L \, E^n \right) \, . \label{equ:ddec}
\end{equation}
Here $D_\kappa (E) $  parameterizes effects due to quantum decoherence and $U_{\alpha i}$ are the elements of
the standard neutrino mixing matrix. Here $n$ carries the energy dependent imprint of a specific model. In the literature, $n=-1,0,2$ have been used (see also
\Ref~\cite{Blennow:2005yk}). In principle, $\Psi$ and $\delta$ can take different values, however, in what
follows we will assume  the same energy dependence for the two parameters.

 From \equ{pdecoh}, we note that quantum decoherence scenario {\it{always}} leads to flavor equipartition
($1:1:1$) as we take the asymptotic limit $L \to \infty$ if both $\Psi,\delta > 0$, whereas  in the limit $\Psi,
\delta \to 0$, we recover the standard oscillation result (\equ{pso}). Physically, the meaning of flavor
equipartition ($1:1:1$) is  that neutrino flavor gets completely randomized due to interaction of system with
 the environment and all the flavors get equally populated.  This  result does not depend on the particular type of
source or choice of mixing parameters. However, as noted before flavor equipartition  was also obtained in the
standard oscillations (without decoherence) specifically for the pion beam source for the best-fit values of
mixing angles. This coincidence is purely accidental and there is no physical reason for this coincidence. In order to infer
effects due to quantum decoherence therefore the pion beam source is rendered useless. However, any of the
anomalous sources (which do not lead to accidental flavor equipartition) are thought to be good candidates for
the study of decoherence vis-a-vis standard oscillations. For instance, it is shown that detection of
 galactic electron anti-neutrino beams at IceCube will lead to a major improvement in sensitivity to
  quantum decoherence effects~\cite{Hooper:2004xr,Hooper:2005jp,PhysRevD.72.065019}. We find such a source at
   low energies or the extreme
high-$B$ region in \figu{hillas}. However, the neutrino energies are typically too low in these cases to have
flavor identification in the neutrino telescopes (the extremely high magnetic field leads to proton synchrotron
losses such that it is difficult to accelerate to high energies).

For a  fixed value of $n$ (which characterizes a particular model), we choose the following  four cases:
\begin{enumerate}
\item $\delta = \Psi = 0$ (standard oscillations)
\item $\delta = \Psi > 0$
\item $\delta=0$ and $\Psi > 0$
\item $\Psi=0$ and $\delta > 0$
\end{enumerate}
Since $n=0$ leads to an energy-independent effect, we do not consider this case. Therefore, we study the
following two cases:
\begin{enumerate}
\item[{(a)} ]  $n=2$: This corresponds to a string inspired model~\cite{Mavromatos:2007hv},
 \\
\begin{equation}
   D_{\kappa} (E) =
\exp \left( - 2 \kappa L E^2 \right) \, .\label{equ:ddecn2}
\end{equation}
Here the quantum decoherence effects are present at high energies. From the exponential factor, setting $2
\kappa L E^2 \sim {\cal O} (1)$ we have decoherence for
\begin{eqnarray}
\kappa^{-1} & \lesssim & 2 \times 10^{44} \,\frac{L}{\rm{Mpc}} \left(\frac{E}{\rm{TeV}}\right)^2 ~{\rm{GeV}} \,.
\end{eqnarray}
This implies that astrophysical neutrinos with  $E= 1$ TeV and $L=1$ Mpc allow us to probe $\kappa \simeq
10^{-44} ~{\mathrm{GeV^{-1}}}$ for this case.
\item [{(b)}]
 $n=-1$: This corresponds to a Lorentz invariant model~\cite{Mavromatos:2007hv},
 \\ \begin{equation}
   D_\kappa (E) =
\exp \left( - 2 \kappa  \frac{L}{E} \right) \, .\label{equ:ddecn2B}
\end{equation}
Here the quantum decoherence effects are present at low energies. The spectral signature is similar to decay
(\equ{pdecay}).
 From the exponential factor, setting $2 \kappa L/E \sim {\cal O}(1)$, we have decoherence for
\begin{eqnarray}
 \kappa^{-1} &\lesssim& 2 \times 10^{35} \,\frac{L}{\rm{Mpc}} \frac{\rm{TeV}}{E} ~{\rm{GeV^{-2}}}\,.
 \label{equ:decohnminusone}
\end{eqnarray}
This implies that astrophysical neutrinos with  $E= 1$ TeV and $L=1$ Mpc allow us to probe $\kappa \simeq
10^{-35}\, {\mathrm{GeV^2}}$ for this case.  We note that even though decay and decoherence are two distinct
physical processes, non-zero value of decoherence paramters $\Psi,\delta$ can be used to set a limit on the
neutrino lifetime, from \equ{decohnminusone} we get $\hat\alpha^{-1} = \tau^0/m \simeq 10^{2} \,
{\mathrm{s\,eV}}$ for $E= 1$ TeV and $L=1$ Mpc.
\end{enumerate}

\begin{figure}[t!]
\begin{center}
\includegraphics[width=0.49\textwidth]{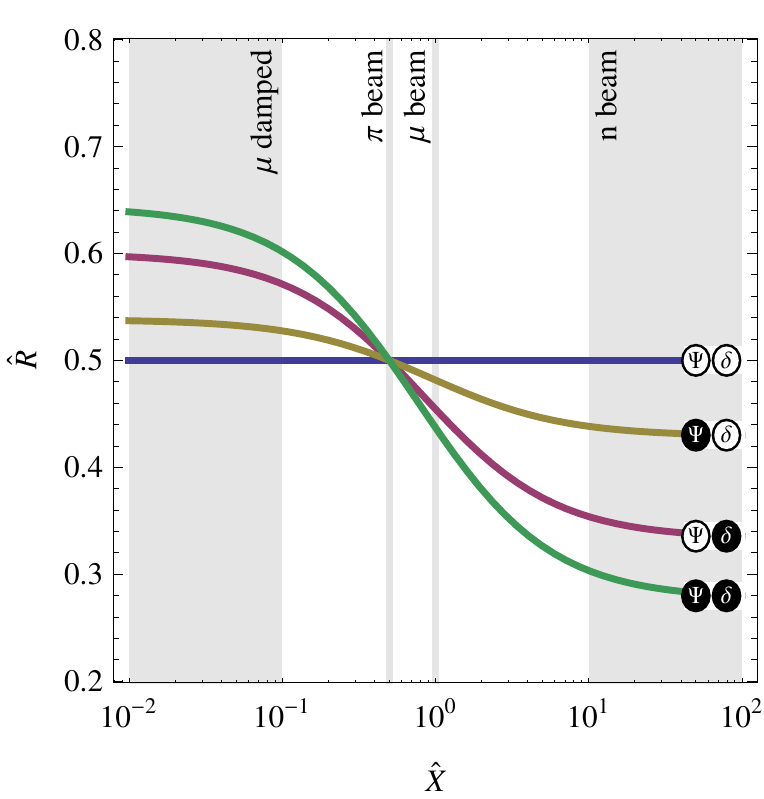}
\includegraphics[width=0.49\textwidth]{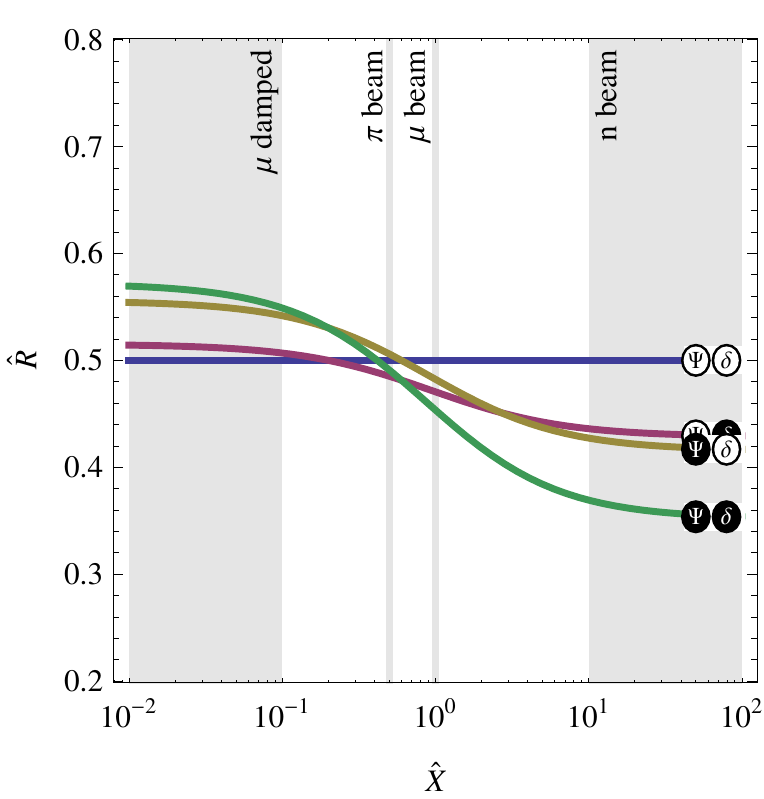}
\end{center}
\mycaption{\label{fig:xdecoh} Flavor ratio $\rhat$ as a function of $\xhat$ for all complete decoherence
scenarios ($L$ large). Black disks refer to coherent parameters ($\kappa=0$), white disks to decoherent
parameters ($\kappa > 0)$. Different sources classes as a function of $\xhat$ are marked. Left figure is for
our best-fit values of the parameters ($\sin^2 \theta_{23} = 0.5$), while the right figure is for $\sin^2 \theta_{23} = 0.4$.}
\end{figure}

We have introduced four scenarios for decoherence, since any decoherence parameter ($\Psi$ or $\delta$) can be
switched on or not. We show in \figu{xdecoh} the flavor ratio $\rhat$ as a function of $\xhat$ for all
   complete decoherence scenarios, \ie, the limit of large enough $L$.
   Black disks refer to coherent parameters ($\kappa=0$), white disks to decoherent parameters
($\kappa > 0)$. As for the decay, we have chosen $\delta=\Psi$ in the case where both parameters are decoherent.
First of all, one can read off from this figure (left panel for maximal atmospheric mixing) that $\delta = \Psi
> 0$ leads to $\rhat=0.5$, irrespective of the value of $\xhat$, since universal flavor mix is a generic
prediction of the decoherence scenario. This can be seen from \equ{pdecoh}. Even more surprising is the fact
that all four scenarios have the same value of $\rhat$ as for complete decoherence case (case 2) for the pion
beam. This is actually due to the conspiracy between the use of best-fit values and the value of $\xhat$ which
leads to vanishing of terms dependent on the mixing matrix elements in \equ{pdecoh} for other three cases,
including that of standard oscillations (see \App~\ref{app:decoh}). If we change one of the mixing angles
($\sin^2 \theta_{23}$), the curves indeed separate out for the pion beam source (see \fig.~\ref{fig:xdecoh},
right panel). The biggest splitting is obtained for the neutron beam, whereas the scenarios can also be
distinguished for the muon beam and muon damped source, in principle. In these cases, the curve $  \Psi=\delta
> 0$ deviates strongest from the standard case $\Psi =\delta =  0$. Note that compared to neutrino decay, all
the different scenarios are relatively close to each other, which means that in practice it may be extremely
difficult to disentangle them.

\begin{figure}[t!]
\begin{center}
\includegraphics[width=\textwidth]{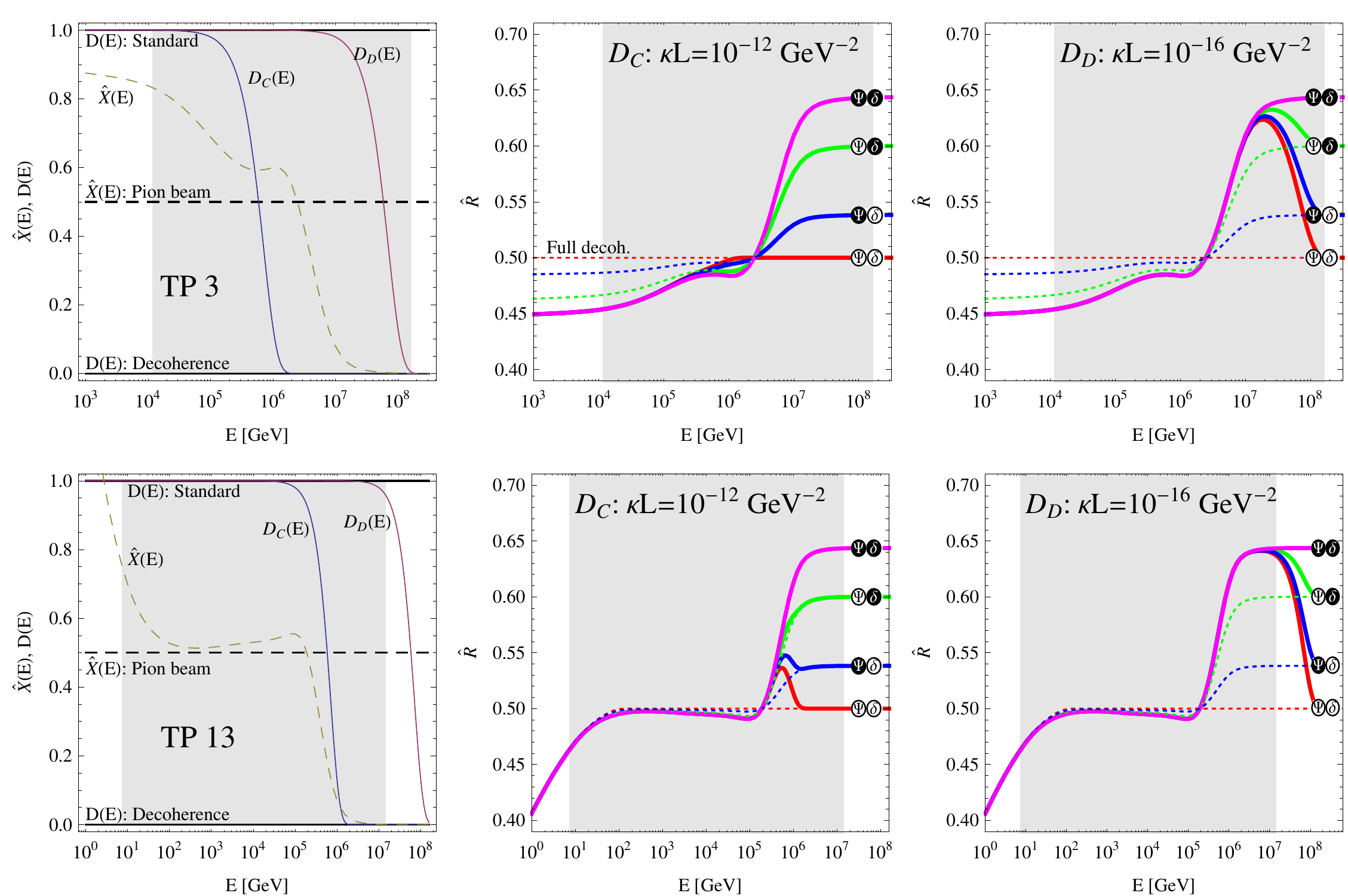}
\end{center}
\mycaption{\label{fig:decohall2} Left panels: Energy dependence of the decoherence coefficients for case (a)
($D_{\kappa} (E) = \exp \left( - 2 \kappa L E^2 \right)$) for $\kappa L=10^{-12} \, \mathrm{GeV}^{-2}$ ($D_C$)
and for  $\kappa L=10^{-16} \, \mathrm{GeV}^{-2}$ ($D_D$), as well as for the function $\xhat(E)$ (dashed curves). The upper row
corresponds to TP~3 from \figu{hillas}  (AGN nuclei), the lower row to TP~13. Middle and right panels: Energy
dependence (solid curves) of the flavor ratio $\hat R$  for the four different decoherence scenarios in case
(a). Black disks refer to coherent parameters ($\kappa=0$), white disks to decoherent parameters ($\kappa > 0)$.
The middle panels are plotted for $\kappa L=10^{-12} \, \mathrm{GeV}^{-2}$  ($D_C$) and the right panels for
$\kappa L=10^{-16} \, \mathrm{GeV}^{-2}$  ($D_D$). The region where the flux is relatively large is shown as
shaded region.   The dotted curves refer to complete decoherence.}
\end{figure}

We show in \figu{decohall2}, the four possible scenarios for case (a) ($n=2$), which are qualitatively different
from   decay   because  here decoherence enters at high energies. This figure is similar to \figu{decayall},
with the white (black) disks referring to decoherence (no decoherence) from the specific parameter. The middle
and right panels correspond to two different values of the
 decoherence parameters~\footnote{For a distance of $L=1$ Mpc, the chosen values for case (a) ($n=2$) correspond to $\kappa \sim
10^{-50}~ {\mathrm{GeV^{-1}}} $ for $D_C$ ($\kappa L =10^{-12} \, \mathrm{GeV}^{-2} $) and $\kappa \sim
10^{-54}~ {\mathrm{GeV^{-1}}}$ for $D_D$ ($\kappa L =10^{-16} \, \mathrm{GeV}^{-2} $) respectively. The chosen
values for case (b) ($n=-1$) correspond to
$\kappa \sim
10^{-32}~ {\mathrm{GeV^2}} $  for $D_E$ ($\kappa L =10^{6} \, \mathrm{GeV} $) and
$\kappa \sim 10^{-30}~ {\mathrm{GeV^2}} $ for $D_F$ ($\kappa L =10^{8} \, \mathrm{GeV} $) respectively.
Specifically, for case (a), that is about $10^{24}$ times more sensitive compared to terrestrial long baseline
experiments, while for case (b) the sensitivity is about  $10^{9}$ times higher (see
\Ref~\cite{Mavromatos:2007hv}). The reason for higher sensitivities is the much longer distances, and, in case
(a), also the higher energies involved.}.
As expected from the energy dependence of   decoherence induced damping, the main effects are present for high
energies, where the scenarios differentiate thanks to the muon-damped behavior of both the test points. In these
cases, they approach the ``full decoherence'' curves (dotted). Comparing $D_C$ (middle panels) and  $D_D$ (right
panels), $D_C$ catches a part of the transition region between the different source types. However, since the
scenarios are quite alike there, this does not help. The decoherence scenario $D_D$, however, leads to effects
practically beyond the peak of the spectrum. Therefore, $\kappa L=10^{-16} \, \mathrm{GeV}^{-2}$ is about the
maximal sensitivity one can achieve. Comparing TP~3 (upper row) with TP~13 (lower row), one can see from the
dotted curves that one could, in principle, also use the range $E \ll 10^6 \, \mathrm{GeV}$ in the upper
 middle
panel for the case of complete decoherence. However, the decoherence effects do not show up at such  low
energies for the chosen parameter values.

\begin{figure}[t!]
\begin{center}
\includegraphics[width=\textwidth]{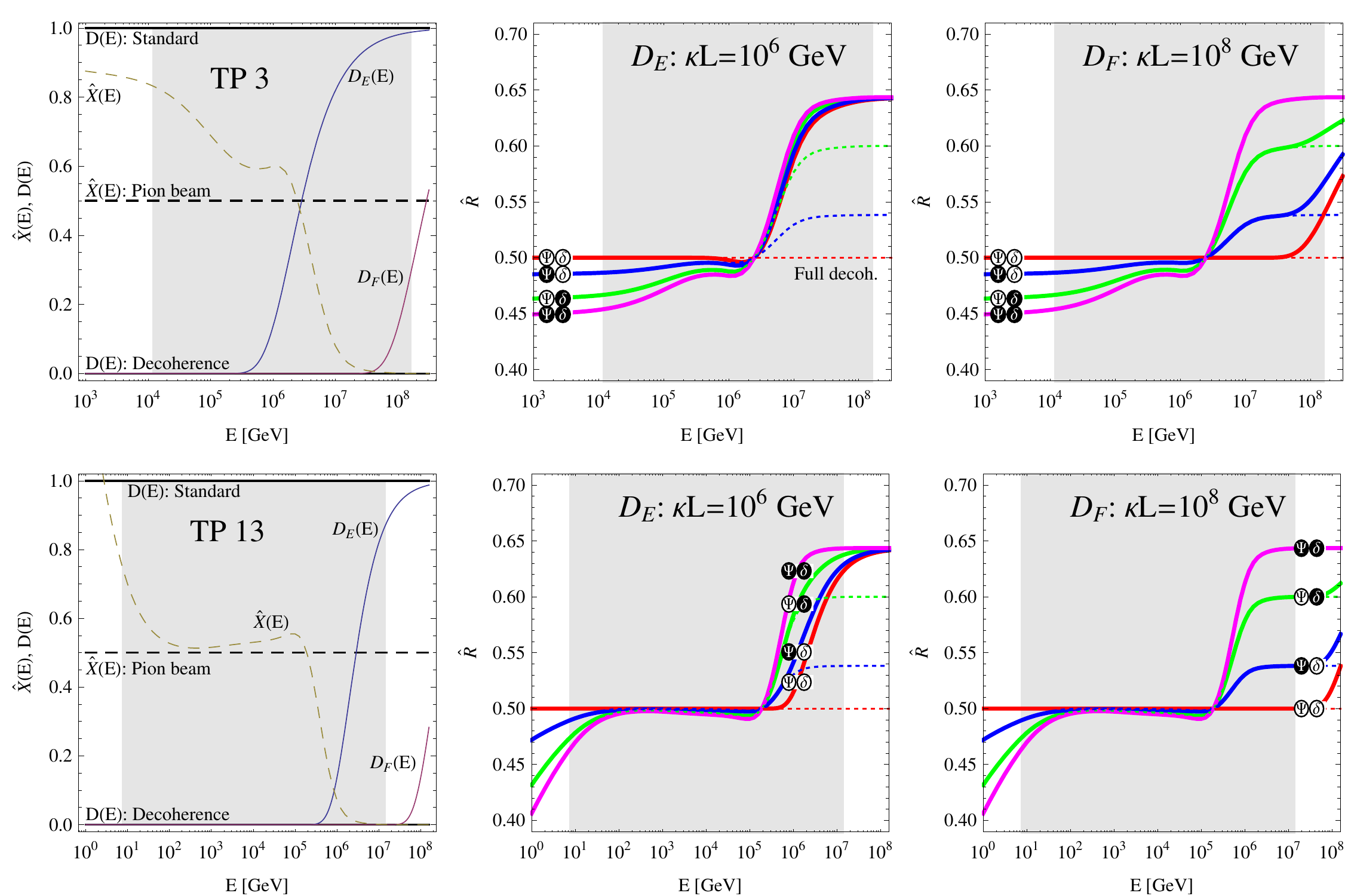}
\end{center}
\mycaption{\label{fig:decohall} Left panels: Energy dependence of the decoherence coefficients for case (b)
($D_\kappa (E) = \exp \left( - 2 \kappa  \frac{L}{E} \right)$) for $\kappa L=10^6 \, \mathrm{GeV}$ ($D_E$) and
for  $\kappa L=10^8 \, \mathrm{GeV}$ ($D_F$), as well as for the function $\xhat(E)$ (dashed curves). The upper row corresponds
to TP~3 from \figu{hillas}  (AGN nuclei), the lower row to TP~13. Middle and right panels: Energy dependence
(solid curves) of the flavor ratio $\hat R$  for the four different decoherence scenarios in case (b). Black
disks refer to coherent parameters ($\kappa=0$), white disks to decoherent parameters ($\kappa > 0)$. The middle
panels are plotted for $\kappa L=10^6 \, \mathrm{GeV}$  ($D_E$) and the right panels for $\kappa L=10^8 \,
\mathrm{GeV}$  ($D_F$). The region where the flux is relatively large is shown as shaded region.   The dotted
curves refer to complete decoherence.}
\end{figure}

In \figu{decohall}, the case (b) ($n=-1$) is shown for comparison. In this case, the energy dependence is
similar to  decay, and the effect of decoherence is present at lower energies. As for decay, it is very
important here that the decoherence effect is large enough to catch a part of the muon-damped range. However,
compared to decay, none of the scenarios are distinguishable at low energies, and one has to rely on the high
energy information. In case (b) (compared to case (a)), the range for the decoherence parameters for which the
scenarios (for $D_E$ and $D_F$) can be distinguished, is small. For example, in the middle panels ($D_E$) the
decoherence parameters are not large enough to produce observable effects in the muon-damped range. In the right
panels ($D_F$), there is some energy range where the scenarios can be distinguished. At high energies, the
muon-damped fixed point is then approached again by all scenarios, which is similar to decay.

\section{
Potential sources on Hillas plot for new physics searches} \label{sec:hillas}

\begin{figure}[t!]
\begin{center}
\includegraphics[width=0.49\textwidth]{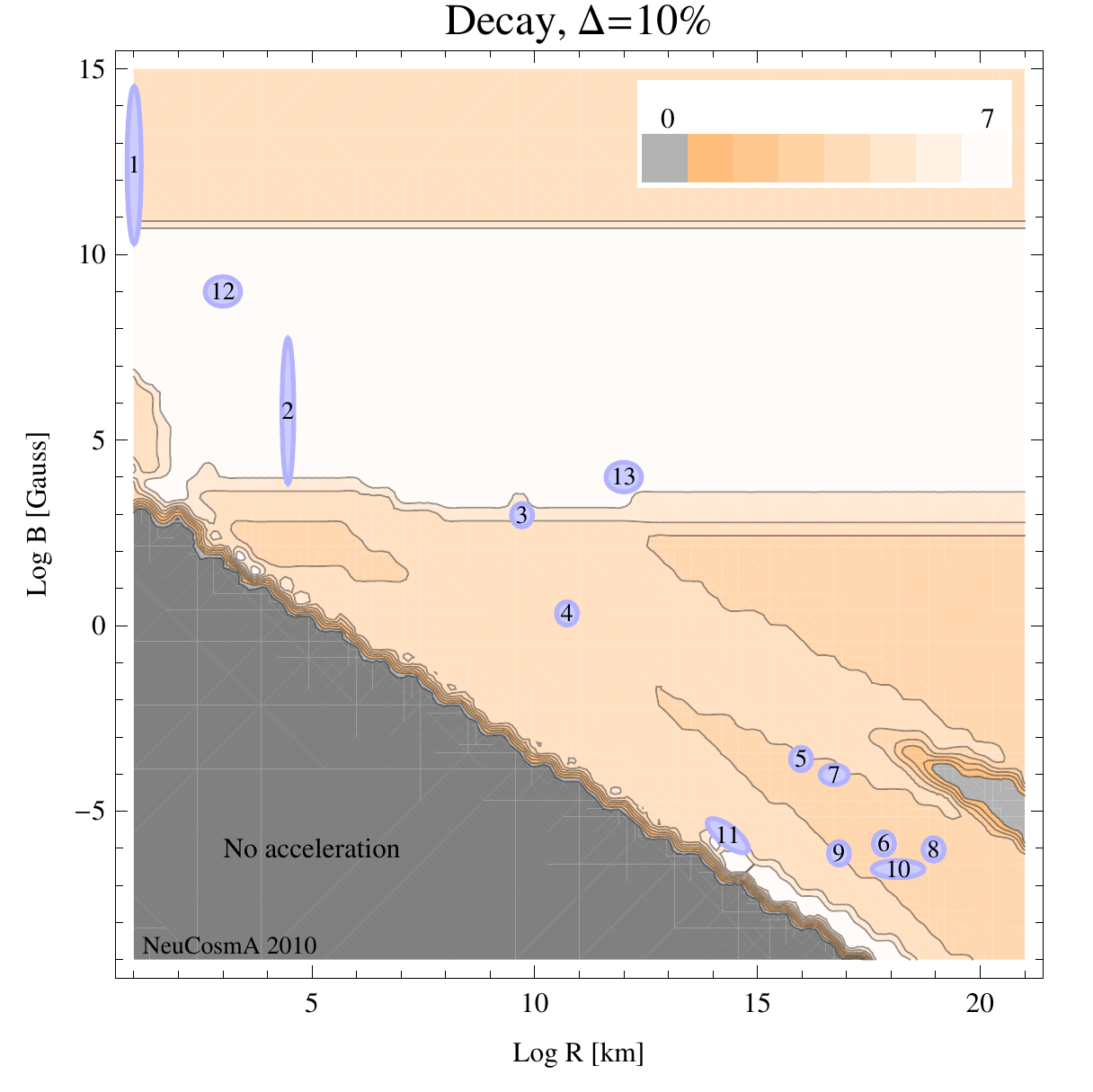}
\includegraphics[width=0.49\textwidth]{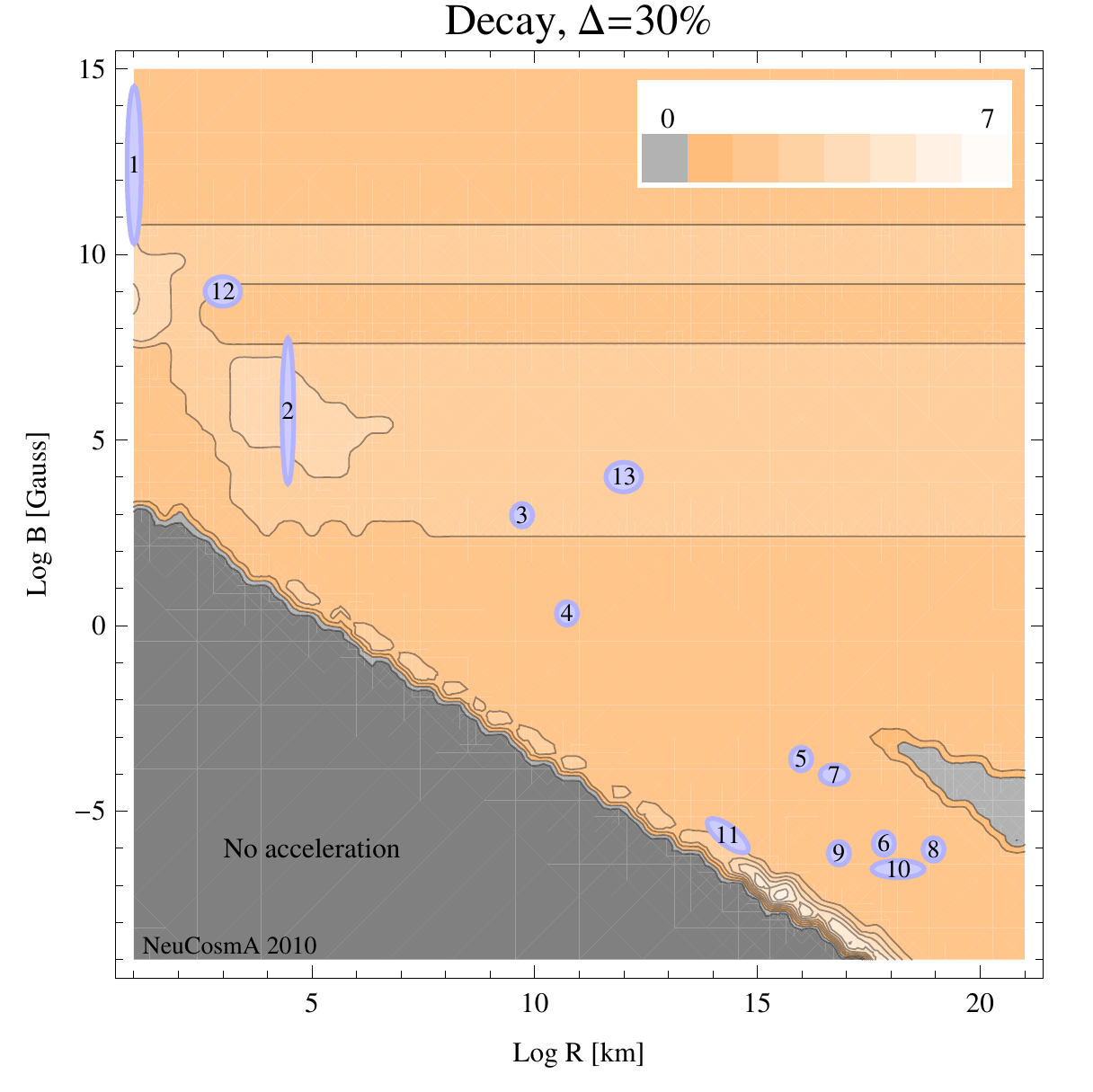}
\end{center}
\mycaption{\label{fig:hillasdecay} Number of decay scenarios (color coding: see legend) which can be uniquely
 determined for the model from \Ref~\cite{Hummer:2010ai} on the Hillas plot. In the left panel, the tolerance
  $\Delta=10\%$, in the right panel, $\Delta=30\%$. Here $\hat \alpha L = 10^8 \, \mathrm{GeV}$ ($D_B$) and
    $\alpha=2$. See main text for details.  }
\end{figure}

Here we discuss which source classes may be especially useful for the purpose of extraction of energy dependent
new physics scenarios. We do not take into account instrumental or mixing parameter uncertainties. We also do not
perform a statistical treatment, but we comment on statistics at the end of this section. We study where to look for possible new physics effects from the theoretical
point of view, \ie, under   ideal set of assumptions.

As method, we count the number of our (seven) decay or (four) decoherence scenarios which we can uniquely
determine for any energy around the spectral peak at a specific tolerance.  That is, for each pair of parameters
$\{R,B\}$, it is tested if the energy dependent flavor ratio can be distinguished from all other scenarios at
least at some energy value (which can be different for different scenarios).  Then the number of such unique
scenarios is counted. Note that compared to \figu{hillas}, where the source is to be found over  at least one
order of magnitude in energy, we also allow for smaller energy windows. In addition, note that this approach is
conservative from the theoretical point of view, because even though a scenario may not be uniquely established,
some other scenarios may be excluded.  For the numerical analysis, we need to impose some numerical uncertainty
$\Delta$ on $\rhat$ (tolerance). For example, for $\Delta=10\%$ and $\rhat=0.5$ for the encountered (true)
scenario, only scenarios with $\rhat < 0.45$ or $\rhat > 0.55$ can be excluded.

In \figu{hillasdecay}, we show the number of decay scenarios (color coding: see legend) which can be uniquely
determined as a function of $R$ and $B$ for $\alpha=2$. In the left panel, the tolerance $\Delta=10\%$, in the
right panel, $\Delta=30\%$. The lighter the color, the more scenarios can, in principle, be uniquely determined.
The (light) spikes at the bottom ($R \sim 10^{15} \, \mathrm{km}$, $B \sim 10^{-7} \, \mathrm{Gauss}$) come from
a contribution from neutron decays. For the (dark) spike  on the right sides ($R \sim 10^{21} \, \mathrm{km}$,
$B \sim 10^{-5} \, \mathrm{Gauss}$), the energies around the spectral peak are already too high, \ie, the
neutrinos from these sources are practically stable. If the tolerance is low enough, obviously all scenarios can
be identified uniquely if the magnetic field is in the range to allow for both the pion beam and the muon damped
part at high energies; \cf,  \figu{decayall}. If only one type of source is dominant, such as for the pion beam
region in \figu{hillas}, not all scenarios can be disentangled. For the tolerance $\Delta=30\%$, of course,
fewer new physics scenarios can be identified. However, it is clear from \figu{hillasdecay}, that the
potentially interesting sources have substantial magnetic fields $10^3 \, \mathrm{Gauss} \lesssim B \lesssim
10^{11} \, \mathrm{Gauss}$, which are, however, not too high to lead to the entire domination of the neutron
beam or muon damped part. For example, for our TP~3 (AGN cores), depending on the tolerance, three to six
scenarios can be uniquely determined, which is already very good.  Other good candidates might be white dwarfs
and gamma-ray bursts (not shown here, since they are not described well within our toy model).

\begin{figure}[t!]
\begin{center}
\includegraphics[width=0.49\textwidth]{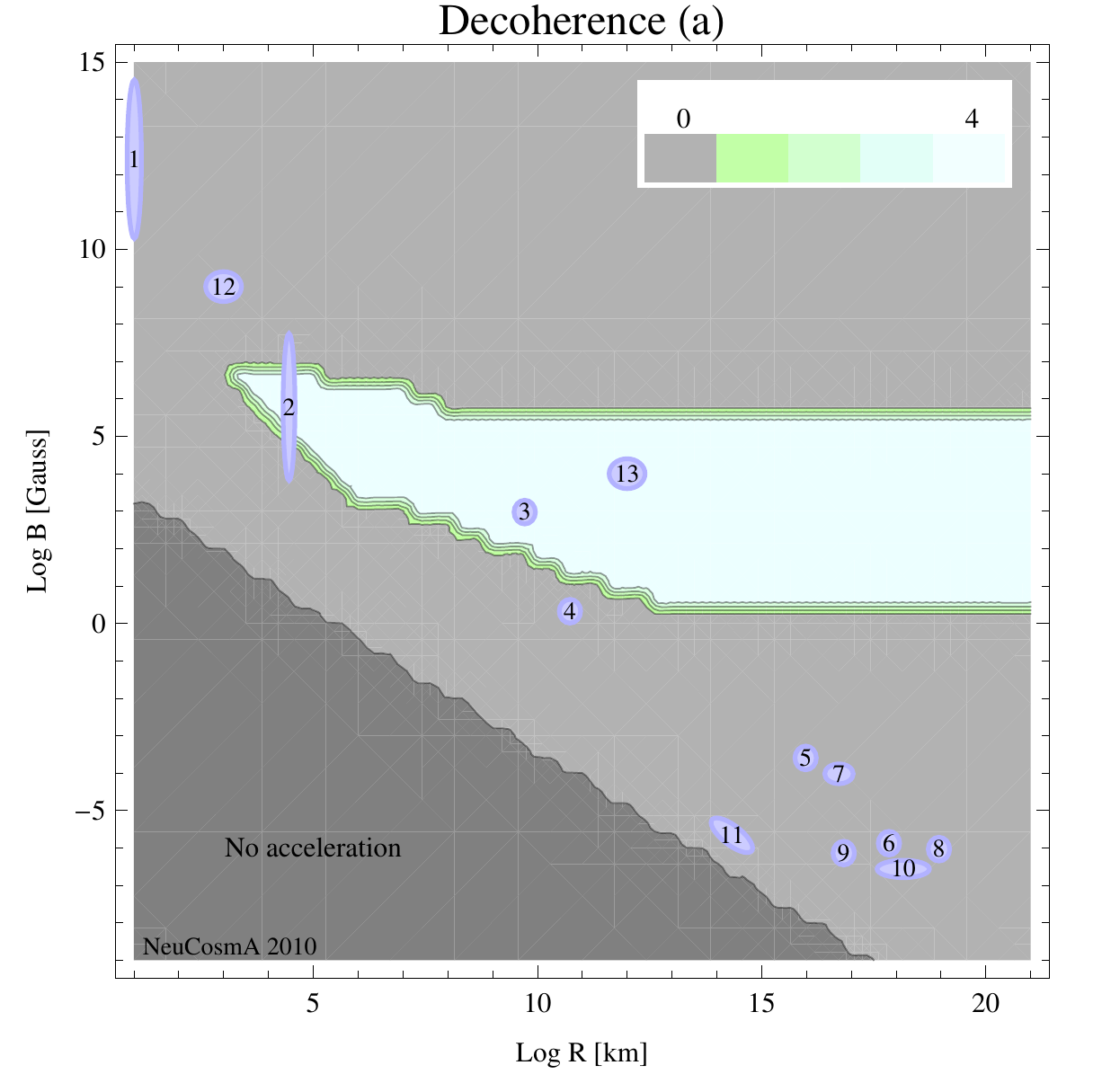}
\includegraphics[width=0.49\textwidth]{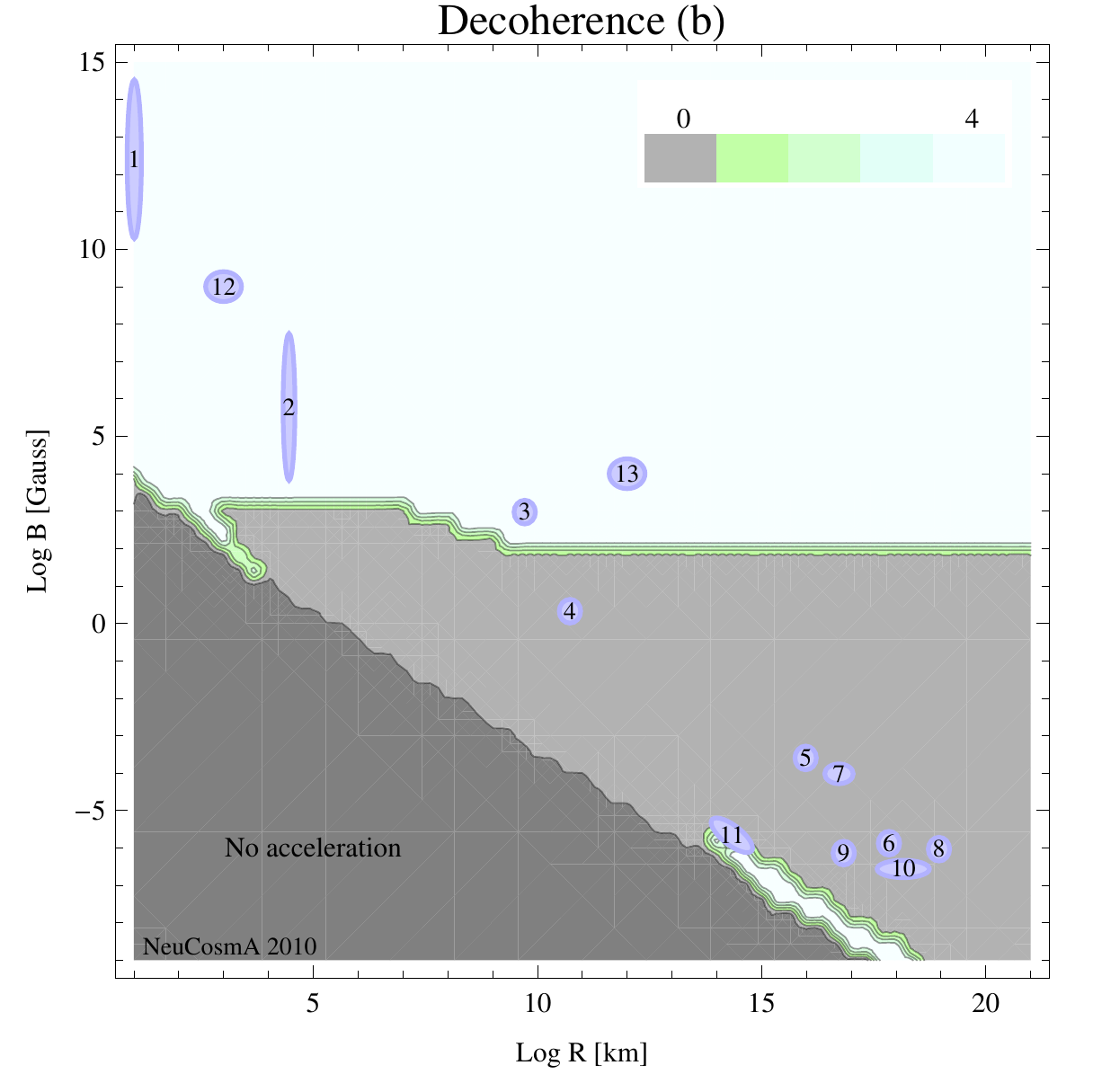}
\end{center}
\mycaption{\label{fig:hillasdecoh} Number of decoherence scenarios (color coding: see legend) which can be
uniquely determined for the model from \Ref~\cite{Hummer:2010ai} on the Hillas plot. In the left panel,
decoherence scenario (a) is shown for $\kappa L=10^{-12} \, \mathrm{GeV}^{-2}$ ($D_C$), in the right panel
decoherence scenario (b) for $\kappa L=10^8 \, \mathrm{GeV}$ ($D_F$). Here $\alpha=2$ and a tolerance on the
flavor ratio of 5\% is assumed. See main text for details.  }
\end{figure}

We show similar  analysis for decoherence scenarios (a) and  (b) in the left and  right panels of
\figu{hillasdecoh}, respectively. Here we have chosen a much smaller values for the tolerance ($\Delta = 5\%$),
since all scenarios are very close to each other. Since the decoherence scenarios are identical for the pion
beam source and split up in equally separated values of $\rhat$ for the muon damped source, there is practically
no qualitative effect of  the number of scenarios: either all or none can be identified.  For both decoherence
scenarios (a) and (b), the pion beam source alone is not sufficient, as we discussed earlier, which means that
substantial magnetic fields are required.  As a qualitative difference between scenarios (a) and (b), (a)
requires high energies for the effect due to decoherence to be present, whereas the signature of (b) is
strongest at low energies. Since proton synchrotron losses in the magnetic field limit the maximal energy in the
considered model, the magnetic field cannot be too strong in case (a), whereas the optimal region for (b) is
basically consistent with the non-pion beam region in \figu{hillas}. Note, however, that for too large $B$, the
neutrino energies may not even be above the cascade threshold in an experiment such as IceCube (unless the source energies are highly Lorentz-boosted). In addition, in
case (a), magnetic fields as low as $10\,{\mathrm{Gauss}}$ may be useful if $R \gtrsim 10^{12} \, \mathrm{km}$,
because then high neutrino energies are obtained while magnetic field effects can be still observed.  In both
cases, TP~2 and TP~3 lie in the optimal region, similar to the decay example above.

In principle, one can also look into the parameter ranges in the individual scenarios which can be constrained.
  We find that decay and decoherence (b) prefer high magnetic fields, where the muon damped (and neutron beam)
   ranges are found at relatively low energies. However, again, note that in this case the event rates are
   surely extremely low, and that the energies may not even be above the cascade threshold in an experiment such
    as IceCube. On the other hand, decoherence (a) scales basically with the maximal energy allowed in the model,
    \ie, lower $B$ and larger $R$ are preferred. However, as discussed, $B$ cannot be too low (lower than about
    1~Gauss), since the effect cannot be seen for the pure pion beam.  Considering the optimal tradeoff between
     high enough neutrino energies and large enough new physics effects, probably the lightest region
     in \figu{hillasdecay}, left, represents the parameter space of the most useful sources for new physics
     effects present at low energies, and the light region in \figu{hillasdecoh} (left), for effects present
     at high energies.  In the range $10^3 \, \mathrm{Gauss} \lesssim B \lesssim 10^6 \, \mathrm{Gauss}$, most
     distinct  scenarios may be accessible.

The discussion in this section has been performed without dedicated statistical study, since it is not  clear
which sources will provide signals and how large these could be.  The number of events $N$ can be estimated by
the convolution of the flux with the exposure (neutrino effective area times time), see, \eg,
\Ref~\cite{IceCube:2010rd} for the effective area of IceCube-40 for different source declinations. In the
absence of backgrounds, which may be used as a first approximation for point sources using the angular
resolution for background suppression, the flux limit scales inversely with $N$. At the 90\% CL, 2.4 events are
still compatible with the non-observation of a flux in about one year of IceCube-40 data. Taking into account
that the effective area of IceCube-86 is about a factor of three to four higher at 100~TeV (see
\Ref~\cite{Karle:2010xx}, Fig.~5) and assuming a total exposure of ten years, about 100 (muon track) events in
total can still be expected in the full-scale experiment if the current bounds are saturated. The effective area
for cascades, which are dominated by $\nu_e$ and $\nu_\tau$  is, however, about a factor of ten lower at
100~TeV~\cite{Karle:2010xx,IcCascade:2011ui}, where it somewhat helps for statistics that we use the fluxes from
both $\nu_e$ and $\nu_\tau$ and do not require an unfolding of the $\nu_e$ events, as, \eg, in
\Ref~\cite{Lipari:2007su}. For flavor equipartition at the detector, this means that for 100 muon tracks only 10
cascades may be expected, and that cascades may constrain the statistics. This however does not apply to all
cases. For instance, if only $m_1$ is stable (see also \Ref~\cite{Lipari:2007su}), $\widehat R \simeq 0.2$
implies that considerably more cascades than muon tracks are expected, up to 50 cascades are compatible with the
current (muon track) bounds, corresponding to a statistical error of $\widehat R$ of only 14\%. Of course, a
unique scenario identification is very difficult, but some scenarios may be excluded then (see, \eg,
\Ref~\cite{Maltoni:2008jr} for a more detailed statistical discussion).  In addition, for detectors such as the
DeepCore array, more muon tracks are fully contained, which means that this discrepancy should be smaller. In
\Ref~\cite{Lipari:2007su}, it was also pointed out that the detector response for muon tracks and showers
depends on the spectral shape, since the energy dependence of these two event classes is different. It may be
regarded as a strength of our approach that we can not only predict the flavor ratio, but also the spectral
shape. Finally, mixing parameter uncertainties have not been taken into account. However, we expect that these
will be reduced by future measurements on the time-scale discussed here to a level at which they are only a
sub-dominant contribution to the total uncertainty, see Fig.~11 in \Ref~\cite{Hummer:2010ai}.

\section{Summary and discussion}
\label{sec:summary}

We have discussed the role of flavor ratio measurements at neutrino telescopes to decipher potential
 new physics effects during the propagation of neutrinos.
While in the literature, flavor ratios at the source are typically considered as constant numbers for particular
source classes (pion beam, muon damped, muon beam, neutron beam), we have incorporated energy dependence in
these  flavor ratios at the source. We assume that these flavor ratios can be predicted well if the astrophysical
parameters of the source are known, such as size of the acceleration region $R$, magnetic field $B$ of the
source, and injection index $\alpha$. In many cases it is possible to estimate these parameters from the
observation of the gamma-ray counterpart, such as by time variability, energy equipartition, and the spectral
shape, respectively.

We have used the HMWY toy model~\cite{Hummer:2010ai} for the flux computation, which can describe relatively
wide parameter regions using as few assumptions as possible. In this case, charged mesons are produced from
photohadronic interactions between protons and the synchrotron photons of co-accelerated electrons/positrons. We
have emphasized that the flavor ratios for a specific energy do not depend on as many assumptions as the
spectral shape. Basically, they are determined by the cooling and escape processes of the secondaries, \ie, the
pions, muons, kaons, and neutrons, whereas any additional cooling or escape processes of the primaries (protons,
electrons, positrons) only affect the spectral shape and maximal energy. Therefore, flavor ratios may serve as
the most robust prediction one may expect from an astrophysical source. However, they are much more difficult to
measure: as minimal assumption, one needs to distinguish between muon tracks and cascades in a neutrino
telescope. The measurement of cascades in this spirit has, for the first time, very recently
 been discussed in \Ref~\cite{IcCascade:2011ui} for IceCube-22.

Apart from the flavor ratio at the source, new physics in the neutrino propagation may also show up as another
energy dependent effect. We have used two examples for such new physics effects, (invisible) neutrino decay and
quantum decoherence, to illustrate the interplay between energy dependent flavor ratios and energy dependent new
physics effects. Neutrino decay characteristically shows up at low energies, where the lifetime of the particles
is shortest because of the smaller Lorentz boost. Quantum decoherence, however, may plausibly be a high energy
effect, at least in certain scenarios discussed in the literature. While it  is useful to have different source
classes as a function of energy in a neutrino source to discriminate different decay scenarios, the considered
quantum decoherence prefers sources other than a pure pion beam source. The reason is that both the standard
case (flavor mixing only) and the quantum decoherence (even partial decoherence) case lead to approximate flavor
equilibration for the pion beam case, which means that they are indistinguishable.

As the final step, we have translated these requirements into our model parameters,  where especially the
magnetic field in the source plays an important role. For high magnetic fields, proton synchrotron losses limit
the maximal energies, which can render high energy ($n=2$) quantum decoherence invisible. In extreme cases, the
neutrino energies may even be too low to pass the neutrino telescope threshold. For too low magnetic fields, the
muon damping is not present in the flavor ratio which means that all decoherence scenarios and some decay
scenarios are practically indistinguishable. Optimal magnetic fields are found to be in the range $10^3 \,
\mathrm{Gauss} \lesssim B \lesssim 10^6 \, \mathrm{Gauss}$, where most effects are observable. Optimal (known)
potential source classes may be AGN cores, white dwarfs, or GRBs, whereas   AGN jets can only discriminate few
effects as they typically perform as pion beams.

We conclude that, although tough to measure, flavor ratios are a very interesting possibility to constrain
particle physics properties using astrophysical sources in parameter ranges which would be otherwise not
accessible. Instead of parameterizing the flavor composition at the source by constant (energy-independent)
numbers, one could take into account the energy dependence of the flavor ratio which may be indirectly obtained
from the multi-messenger connection. In many new physics scenarios, it is exactly the  combination of this
energy dependent flavor composition and the energy dependent new physics effect which may lead to new
interesting observations or constraints -- even with a single source, if enough statistics can be obtained.
Although a high statistics measurement may not be expected from IceCube, future neutrino telescopes may be
better optimized for cascade detection.

\subsubsection*{Acknowledgments}

WW acknowledges support from Deutsche Forschungsgemeinschaft, Emmy Noether grant WI 2639/2-1. PM would like to
thank the Institut f\"{u}r Theoretische Physik und Astrophysik at Universit\"{a}t W\"{u}rzburg for their kind
hospitality during her visit in July 2010.

\appendix

\section{Three flavor neutrino oscillation probability in presence of decoherence}
\label{app:decoh}

Here we describe a phenomenological ``bottom-up" framework to describe deviations from standard neutrino
oscillation formalism in presence of decoherence irrespective of the details of the underlying theory (\eg,
quantum gravity). Such a quantum system  is described in terms of density matrices and the Liouville
 equation for the neutrino flavor density matrix
$\rho$~\cite{Ellis1984381,Banks1984125,Liu:1993ji,Benatti:2000ph,Ohlsson:2000mj,Morgan:2004vv,Hooper:2005jp,PhysRevD.72.065019}
(for any number of flavors) is modified to
\begin{equation}
\dot \rho = -i \left[H,\,\rho\right] +  {\cal D} [\rho] \label{equ:eom}
\end{equation}
where $H$ is the Hamiltonian  and $\dot \rho$ implies differentiation with respect to time. The first term is
responsible for unitary evolution while the term $ {\cal D} [\rho]$ contains effects due to non-unitary
evolution. A commonly used form for ${\cal D} [\rho]$ was  given by Lindblad using quantum dynamical
semi-groups~\cite{lindblad}.
 For an $N-$level quantum system, it is possible to expand all the operators in the $SU (N)$
  Hermitian basis
  \begin{eqnarray}
\rho &=& \frac{1}{2} \left[p_\mu \lambda_\mu \right] = \frac{1}{2} \left[p_0 I +  p_i  \lambda_i \right] \nonumber\\
H &=& \frac{1}{2} \left[h_\mu  \lambda_\mu \right]  = \frac{1}{2} \left[h_0 I +  h_i \lambda_i \right] \nonumber\\
{\cal D} [\rho] &=& \frac{1}{2} \left[ \lambda_\mu d_{\mu\nu}\rho_\nu \right]
\end{eqnarray}
  which leads to the equation of motion in the component form
\begin{equation}
\dot p_\mu = (h_{\mu \nu} + d_{\mu\nu}) p_\nu \label{equ:eomcomp}
\end{equation}
where the subscripts $\mu,\nu$ depend upon the number of flavors, for three flavors, $\mu,\nu=0,1,...,8$. The
matrix elements $h_{\mu\nu}$ are usually fixed from the form of the Hamiltonian while the elements of this
matrix $d_{\mu\nu}$ can be fixed by assuming that the laws of thermodynamics hold.
 If one imposes the requirement of the monotonic increase of von-Neumann entropy ($S=-{\rm {Tr}}[\rho \ln \rho]$) which leads to hermiticity of operators,
conservation of average value of energy, conservation of probability etc, we can further constrain the elements,
$d_{\mu\nu}$. For $N=3$ case, the basis is given by eight Gell-Mann $SU(3)$ matrices
 $\{\lambda_i\}$ along with the Identity $I_3$ matrix.  %
 We have the following matrix equation (from \equ{eomcomp}) with eleven decoherence parameters as in \Ref~\cite{Hooper:2005jp},
\begin{eqnarray} \frac{d}{dt}\begin{pmatrix}
   p_0 \\
     p_1 \\
     p_2 \\
     p_3 \\ p_4\\p_5\\p_6\\p_7\\ p_8
\end{pmatrix} &=& \left[
  \begin{pmatrix}
  0 & 0 & 0 & 0 & 0
&0&0&0&0\\
   0  & A & B+\omega_{21} & 0&0
&0&0&0&0\\
    0 & B- \omega_{21} & \Lambda & 0&0
&0&0&0&0\\
     0 &0 &0&\Psi & 0
&0&0&0&0\\
     0 &0 &0&0& x
&y+\omega_{31}&0&0&0\\
     0 &0 &0&0&y-\omega_{31}
&z&0&0&0\\
     0 &0 &0&0&0&0
&a&b+\omega_{32}&0\\
     0 &0 &0&0&0
&0&b-\omega_{32}&\alpha&0\\
     0 &0 &0&0&0
&0&0&0& \delta
\end{pmatrix} \right]\begin{pmatrix}
   p_0 \\
     p_1 \\
     p_2 \\
     p_3 \\ p_4\\p_5\\p_6\\p_7\\ p_8
\end{pmatrix}
\label{equ:fullmateqn}~, \nonumber\\
\end{eqnarray}
where $\omega_{ij}=\delta m_{ij}^2/4E$  with $\delta m_{ij}^2=m_i^2-m_j^2$ is the standard oscillation
 term.  $p_0 =2/3 $ by requiring $Tr[\rho]=1$. We impose ${\rm{Tr}}[\rho]=1$ and ${\rm{Tr}}[\dot \rho] = 0$ which leads
to conservation of probability.  This leads to first row and column of the total matrix in above equation to be
zero. Trace is preserved during evolution and hence $p_0$ remains constant. The remaining eight components of
$\rho(t)$ can be obtained by solving  the
  set of eight coupled equations subject to the initial condition
\begin{eqnarray}
\rho_{\nu_\alpha} (0)  &=&   \begin{pmatrix} U_{\alpha 1} ^2  & U_{\alpha 1} U_{\alpha 2}
   & U_{\alpha 1} U_{\alpha 3} \\
 U_{\alpha 2} U_{\alpha 1}  &   U_{\alpha 2} ^2 & U_{\alpha 2} U_{\alpha 3}  \\
U_{\alpha 3} U_{\alpha 1}  &  U_{\alpha 3} U_{\alpha 2} &   U_{\alpha 3}^2 \end{pmatrix} \label{equ:rhoal}~,
\end{eqnarray}
 and noting that the density matrix is given by
\begin{eqnarray}
\rho (t) &=&  \frac{1}{2} \begin{pmatrix} \frac{2}{3} + p_3 +\frac{p_8}{\sqrt 3} &
  p_1- i p_2 & p_4 -i p_5 \\
 p_1+i p _2 & \frac{2}{3} - p_3 + \frac{p_8}{\sqrt 3}  & p_6 -i p_7\\
p_4+i p_5 & p_6 + i p_7& \frac{2}{3}  -\frac{2 p_8}{\sqrt 3} \end{pmatrix}~.\label{equ:rhogen}
\end{eqnarray}
The diagonal elements of $\rho (t)$ are referred to as populations while the off-diagonal elements
  as coherences. The phase information is contained in the coherences of the density matrix.
 Let us define the $8 \times 8$ block in \equ{fullmateqn} connecting the components
  $p_i (i\neq 0)$ by ${\cal L}$
\begin{eqnarray}  {\cal L} &=&
  \begin{pmatrix}
   A & B+\omega_{21} & 0&0
&0&0&0&0\\
   B- \omega_{21} & \Lambda & 0&0
&0&0&0&0\\
   0 &0&\Psi & 0
&0&0&0&0\\
   0 &0&0& x
&y+\omega_{31}&0&0&0\\
    0 &0&0&y-\omega_{31}
&z&0&0&0\\
    0 &0&0&0&0
&a&b+\omega_{32}&0\\
    0 &0&0&0
&0&b-\omega_{32}&\alpha&0\\
    0 &0&0&0
&0&0&0& \delta
\end{pmatrix}
\end{eqnarray}
   and denote ${\cal M} = e^{-2 {\cal
L} t}$ and then write the solutions $p_i (t)$ in terms of the elements of exponentiated matrix ${\cal M}$,
\begin{eqnarray}
p_1 (t) &=& p_1 (0) {\cal M}_{11} + p_2 (0) {\cal M}_{12} + \ldots + p_8 (0) {\cal M}_{18} \nonumber \\
p_2 (t) &=& p_1 (0) {\cal M}_{21} + p_2 (0) {\cal M}_{22} + \ldots + p_8 (0) {\cal M}_{28} \nonumber \\
\vdots \nonumber\\ p_8(t) &=& p_1 (0) {\cal M}_{81} + p_2 (0) {\cal M}_{82} + \ldots + p_8(0) {\cal M}_{88}~,
\label{equ:pasm}
\end{eqnarray}
where  $p_{i} (0)$ are the components of the initial density matrix given in \equ{rhoal} and \equ{rhogen}.

Finally the neutrino oscillation probability $P_{\alpha \beta}$ can be computed using
\begin{eqnarray}
P_{\alpha \beta} (t) &=& {\mathrm{Tr}} [\rho_{\nu_\alpha} (t) \, \rho_{\nu_\beta} (0)]~,
\end{eqnarray}
where $\rho_{\nu_\beta} (0)$ is the ``pure" neutrino density matrix corresponding to flavor $\nu_\beta$ at $t=0$
and $\rho_{\nu_\alpha} (t)$ is the density matrix at $t$ for flavor $\nu_\alpha$.

For obtaining general expression of probability for astrophysical neutrinos, we need to average over $\sin$ and
$\cos$ terms, which gives us
\begin{eqnarray}
 P_{\alpha\beta}  = \frac{1}{3} +
   \frac{1}{2} (U_{\alpha 1}^2 - U_{\alpha 2}^2)(U_{\beta 1}^2-U_{\beta 2}^2)
 D_{\Psi}  +
 \frac{1}{6}  (U_{\alpha 1}^2+U_{\alpha 2}^2- 2
 U_{\alpha 3}^2)(U_{\beta 1}^2+U_{\beta 2}^2 - 2 U_{\beta 3}^2)
  D_{\delta}
 \, ,
 \label{equ:pdecohfinapp}
\end{eqnarray}
where,  $D_{\Psi} =\exp\{-2 \Psi t\}$ and $D_{\delta}=\exp\{-2\delta t\}$ are the decoherence-induced damping
factors. Thus out of the eleven decoherence parameters appearing in \equ{fullmateqn}, only two ($\Psi$ and
$\delta$) appear in the final expression which are the $\lambda_3$ and $\lambda_8$ components in the decoherence
matrix. If we look at the form of $\rho (t)$, this implies that the coherences vanish for astrophysical
neutrinos.
  This means that
the rest of the physically allowed decoherence parameters are inaccessible by astrophysical neutrinos. If any of
the phase information could be retained (which is what happens in atmospheric neutrino case) then the other
decoherence parameters (corresponding to coherences in the $\rho (t)$) will also appear in the
probability~\cite{Hooper:2005jp}. Let us now discuss the special cases in the large $t$ limit:
\begin{itemize}
\item $ \Psi \neq 0, \delta = 0$: \begin{eqnarray}
 P_{\alpha\beta}  = \frac{1}{3} +
 \frac{1}{6}  (U_{\alpha 1}^2+U_{\alpha 2}^2- 2
 U_{\alpha 3}^2)(U_{\beta 1}^2+U_{\beta 2}^2 - 2 U_{\beta 3}^2)
 \, ,
 \label{equ:pdecohfinapp1}
\end{eqnarray}
\item $ \Psi=0, \delta \neq 0$: \begin{eqnarray}
 P_{\alpha\beta}  = \frac{1}{3} +
 \frac{1}{2} (U_{\alpha 1}^2 - U_{\alpha 2}^2)(U_{\beta 1}^2-U_{\beta 2}^2)
 \, ,
 \label{equ:pdecohfinapp2}
\end{eqnarray}
\item $\Psi = \delta = 0$: \begin{eqnarray}
 P_{\alpha\beta}  =
   \sum\limits_{i=1}^{3}  |U_{\beta i}|^2 \, |U_{\alpha i}|^2 \,
\, .
 \label{equ:pdecohfinapp3}
\end{eqnarray}
\item $ \Psi =\delta \neq 0$: \begin{eqnarray}
 P_{\alpha\beta}  = \frac{1}{3}
 \, ,
 \label{equ:pdecohfinapp4}
\end{eqnarray}
\end{itemize}
We note that only when both $\Psi$ and $\delta$ are non-zero, one gets the equilibrium steady state value for
the probability (\equ{pdecohfinapp4})~\cite{Rajagopal1998237}. This means that a certain combination of non-zero
parameters are responsible for obtaining equilibrium steady state answer or complete irreversibility. Violation
of conservation of energy in the neutrino system is necessary but not a sufficient condition to ensure this.  If
one of the two parameters ($\Psi,\delta$) are zero, the probability depends upon the mixing matrix elements in
general.  If both the parameters ($\Psi,\delta$) are zero, we recover standard oscillation result. We can refer
to these three cases as achieving some sort of a steady state out of equilibrium.


\end{document}